\documentclass[12pt,preprint]{aastex} 

\received{2001 November 21} 
\begin{document} 

\title{Chandra Observations of the Pleiades Open Cluster: X-ray Emission from 
Late-B to Early-F Type Binaries} 

\author{Kathryne J. Daniel\altaffilmark{1,2}, Jeffrey L. Linsky\altaffilmark{2}, and Marc Gagn\'e\altaffilmark{1}} 
\email{\small kdaniel@wcupa.edu, jlinsky@jila.colorado.edu, mgagne@wcupa.edu} 

\altaffiltext{1}{Department of Geology and Astronomy, West Chester University, West Chester, PA 19383} 
\altaffiltext{2}{JILA, University of Colorado and NIST, Boulder, CO 80309-0440}

\begin{abstract} 

We present the analysis of a 38.4~ks and a 23.6~ks observation of the core of 
the Pleiades open cluster.  The Advanced CCD Imaging Spectrometer on 
board the {\it Chandra} X-ray Observatory detected 99 X-ray sources in a 
$17\arcmin\times17\arcmin$ region, including 18 of 23 Pleiades members. 
Five candidate Pleiades members have also been detected, confirming their 
cluster membership.  Fifty-seven sources have no optical or near-infrared 
counterparts to limiting magnitudes V$=22.5$ and J$=14.5$.  The unidentified 
X-ray sources are probably background AGN and not stars.  The {\it Chandra} 
field of view contains seven intermediate mass cluster members.  Five of these, 
HII~980 (B6 + G), HII~956 (A7 + F6), HII~1284 (A9 + K), HII~1338 (F3 + F6), and 
HII~1122 (F4 + K), are detected in this study.  All but HII~1284 have high 
X-ray luminosity and soft X-ray spectra. HII~1284 has X-ray properties 
comparable to non-flaring K-type stars.  Since all five stars are visual or 
spectroscopic binaries with X-ray properties similar to F--G stars, 
the late-type binary companions are probably producing the observed coronal 
X-ray emission.  Strengthening this conclusion is the nondetection by 
{\it Chandra} of two A stars, HII~1362 (A7, no known companion) and HII~1375 
(A0 + A~SB) with X-ray luminosity upper limits 27--54 times smaller than 
HII~980 and HII~956, the B6--A7 stars with cooler companions. 
Despite the low number statistics, the {\it Chandra} data appear to confirm 
the expectation that late-B and A stars are not strong intrinsic X-ray sources. 
The ACIS spectra and hardness ratios suggest a gradual increase in coronal 
temperature with decreasing mass from F4 to K.  M stars appear to have somewhat 
cooler coronae than active K stars. 

\end{abstract} 



\section{Introduction} 

Magnetic dynamo theories predict that an interaction between differential 
rotation and sub-photospheric convection creates large photospheric magnetic 
fields \citep{Par55}.  These magnetic fields interact in stellar coronae to 
produce X-ray emission and flares \citep{Pal81}.  Because an outer convective 
envelope is essential for producing dynamo driven magnetic activity, stars with 
radiative outer envelopes (hotter than $\sim$A2) \citep{Boh84} and stars with 
thin outer convective envelopes (late-A to early-F) do not show a strong 
correlation between X-ray emission and rotation as would be expected for 
coronal magnetic activity \citep{Wal83}.  Various theories, such as 
heating through the dissipation of acoustic and MHD waves 
\citep{Ste81,NU90}, have been suggested to explain X-ray activity from late-A 
to early-F type stars (see discussion in \S4.2).  

One of the early surprises from the {\it Einstein} mission was the detection 
of strong X-ray emission from O stars, with X-ray luminosity highly correlated 
with bolometric luminosity \citep{Har79}.  \citet{LW80} suggested that shock 
heated clumps could form in the radiatively driven winds of OB stars, possibly 
leading to X-ray emission.  More recent instability-driven wind shock models 
\citep*{Owo88, MC89, Hil93, Fel95, Coh96, OC99} have been able to explain the 
X-ray properties of many O and early-B stars with sufficiently high mass-loss 
rates. 

The two leading explanations for X-ray emission, instability-driven wind shocks 
and dynamo driven magnetic activity, probably cannot produce strong X-rays for 
stars in the approximate spectral range B3--early-F \citep{Sta94}.  
Nonetheless, moderate X-ray emission from late-B to early-F stars has been 
reported for nearly twenty years 
\citep*[e.g.,][and references therein]{Cai94,Hue01}.  For 
example, in {\it Einstein} \citep{CH85,Mic90}, {\it ROSAT} 
\citep*{Sta94,Gag95,Mic96,Mic99}, and {\it Chandra} \citep[][, hereafter Paper
I]{Kri01} studies of 
the Pleiades, some fraction (10--50\%) of late-B to early-F stars were 
detected.  Two explanations have been suggested: the X-rays are produced in 
the coronae of unseen late-type companions 
\citep[as first suggested by][]{Gol83} or these stars are intrinsic 
X-ray emitters.  The late-type companion hypothesis is difficult to disprove 
because close binaries are not resolved in X-ray images and spectroscopic 
binaries are not resolved with high-resolution X-ray grating spectra. 

Several studies of the X-ray emission from B- and F-type stars have examined 
the late-type companion hypothesis in detail.  In a {\it ROSAT} study of A-type 
stars, \citet{Pan99} could not conclusively identify the source of the X-ray 
emission. \citet{Mic99} concluded that all the B-type stars detected in the 
Pleiades are known or possible binaries, thereby supporting the late-type 
companion hypothesis.  Similarly, in an {\it Einstein} survey of the Pleiades, 
\citet{CH85} concluded that the X-ray emission from the $\sim$A6 star, 
HII~1384, could be attributed to a late-type companion.  \citet*{Sim95} used 
archival {\it ROSAT} data to study ten A-type stars, of which five are 
associated with known late-type companions.  In this study, \citet{Sim95} 
concluded that the remaining five must have late-type companions.  
\citet{Sta94} performed a {\it ROSAT} survey of the Pleiades showing 
inconclusive results on this subject.  Finally, in {\it Einstein} and 
{\it ROSAT} studies of B6-A3 stars in the Orion Nebula, \citet{CZ89} and 
\citet{Sta94} proposed that the X-ray emission from these stars could be 
attributed to T-Tauri stars.  As \citet{CZ89} also suggested, the current 
models for X-ray emission may need to be revised, as none of these studies 
could prove or disprove the late-type companion hypothesis. 

Several studies have suggested that late-B to early-F stars are intrinsic 
X-ray emitters.  In studies by \citet{Mic96}, 
\citet{Sta94}, and \citet{CH85}, the A star, HII~1384 (see above), exhibits 
steady, intense X-ray emission.  There is no known companion to this star, as  
noted by \citet{Mic90,Mic96}.  An analysis by \citet{BS94}, using data from the 
{\it ROSAT} all-sky survey of B stars with known late- or early-type 
companions, showed that most of the X-ray flux from these systems was emitted 
by the early-type stars and not late-type companions.  A similar examination of 
visual binaries was performed by \citet{Sch93} using {\it ROSAT} HRI pointing 
data.  Here again, they concluded that the dominant X-ray emission was from 
late B-type stars. \citet{Sch97} analyzed observations of A- through G-type 
stars from the {\it ROSAT} all-sky survey.  He consistently found X-ray 
emission from stars later than $\sim$A7 and concluded that shallow convection 
zones are sufficient to produce X-ray emission.  

The Pleiades is the prototypical young open cluster.  It contains a co-eval 
($\sim100$~Myr old) population of early- and late-type stars at approximately 
the same distance \citep[$\sim127$~pc,][]{SN01,Sta94} with constant reddening 
($E(B-V)\approx 0.03$) in all but the northwest corner of the cluster 
\citep{Sta94}.  Pleiades membership probabilities have been determined in a 
number of studies using lithium abundances and proper motion data. Recent 
radial velocity studies of the Pleiades \citep{Liu91,Rab98} combined 
with Hipparcos and Tycho studies of visual binaries \citep{Dom00,Hog00} have 
determined the duplicity of all Pleaides stars hotter than K0.  In this paper, 
we use a pair of deep {\it Chandra} ACIS-I observations centered on the core of 
the Pleiades cluster to further examine the X-ray characteristics of early- and 
late-type cluster members. 

\section{Data Acquisition and Reduction} 

{\it Chandra} observed the core of the Pleiades open cluster for a total of 
56.4~ks. The data consist of a 34.8~ks exposure using the Advanced CCD 
Imaging Spectrometer (ACIS) detector acquired on 1999 September 18 
\citep{Kri01} and a 21.2~ks exposure obtained on 2000 March 20.  
{\it Chandra}'s high-resolution mirror assembly and ACIS are described in 
``The Chandra Proposers' Observatory Guide."  The ACIS-I CCDs (I0-I3) and two 
CCDs on the ACIS-S array (S2-S3) continuously collected 3.24~s exposures with 
$\sim$40~ms read-out intervals.  The I3 aim-points were at (J2000) 
$\alpha = 3^{\rm h}46^{\rm m}45{\stackrel{\rm s}{\textstyle{.}}}6$, 
$\delta = +24^{\circ}04\arcmin31\arcsec$ and 
$\alpha = 3^{\rm h}46^{\rm m}48{\stackrel{\rm s}{\textstyle{.}}}7$, 
$\delta = +24^{\circ}04\arcmin57\arcsec$ 
for the 1999 and 2000 observations, respectively.  

\subsection{Data Reduction} 

We reduced the data using methods available through version 1.1.3 of 
{\it Chandra Interactive Analysis of Observations} (CIAO) software, beginning 
with the second level event list.\footnote{An ``event list" is a file 
cataloging each photon's position at impact with {\it Chandra}'s CCD camera, 
time of arrival and the energy.  The {\it Chandra} X-ray Center performs 
preliminary reduction processes on the data resulting in a ``second level" 
event list.}  Events with standard {\it ASCA} grades (0, 2, 3, 4, and 6) were 
retained.  Several hot columns were visible in each observation that had not 
been removed during the original reduction process.  We extracted 14 ``hot 
columns" per observation and six additional columns per CCD near node 
boundaries.  Subtracting these columns from the original 6144 columns resulted 
in a $\sim0.8\%$ data loss.  We used the aspect histograms and the incident 
spectrum of a typical Pleiades X-ray source to generate energy-corrected 
exposure maps.  From these we produced a map of the effective exposure time 
over both observations.  We used this final exposure map to calculate the count 
rate for each source. 

ACIS event lists often contain ``flaring pixels": 2-7 localized 
high-energy events in consecutive read-out frames.  Using Takamitsu Miyaji's 
C-program, FLAGFLARE,\footnote{FLAGFLARE is available at 
http://www.astro.psu.edu/xray/acis/recipes/clean.html.} all events with 
flare values $<1$ were removed.  \citet{TG} show that $\sim5\%$ of the counts 
from real events may be removed by using FLAGFLARE. 

Finally, the front-side illuminated CCDs were damaged by radiation early in the 
mission.  The consequent charge transfer inefficiency (CTI) affects the gain 
and energy resolution \citep{Tow00} of ACIS.  We corrected the event lists 
for each observation using software provided by the {\it Chandra} X-ray Center. 

\subsection{Source Extraction} 

For each dataset we generated three $1400\times1400$ pixel images: with 
$0.5\arcsec$, $1\arcsec$ and $2\arcsec$ pixels.  The ACIS-S CCDs were far 
off-axis for these observations.  As a result of vignetting and the large 
point-spread function (PSF), no new sources were detected; the ACIS-S data are 
not presented in this paper.  The first and second epoch images were registered 
with respect to the 1999 September images.  The resulting co-added images 
were sharp and showed no signs of systematic astrometric errors. 
Figure~1 shows a $1400\arcsec\times1400\arcsec$ image with $1\arcsec$ pixels 
of the Pleiades with a logarithmic stretch. 

Using these co-added images and exposure maps, we detected 99 {\it Chandra} 
sources by applying CIAO's wavelet-based algorithm tool, {\em wavdetect}.  {\em 
Wavdetect} was applied to each image at scales 2,~4,~and~8, where the scaling 
factors are defined as wavelet radii in pixels.  Using a false alarm 
probability of $10^{-6}$, we should detect $<2$ spurious sources per image.  
We acquired the greatest coordinate accuracy for each source detection by using 
scaling factors in accordance with the PSF, and therefore off-axis angle. 

The individual source lists from 1999 September and 2000 March were visually 
compared to the co-added source list.  We found that twenty-three new sources 
appeared as a result of the co-added image's increased S/N.  An additional 
twenty-two wavdetect detections were clearly artifacts of the detection 
algorithm: these detections were lined up along the edge of the FOV, having 
been detected because the counts inside the source region were significantly 
higher than the background outside the FOV.  None of these twenty-two positions 
corresponded to known optical stars.  In three cases, a 3--4 count source with 
no optical candidate was detected in one of the two individual exposures, but 
not in the co-added image.  These three were eliminated because no other other 
3--4 count sources were detected.  We removed a five count detection because it 
was over $5\arcmin$ off-axis (where the PSF is larger) and no sources with 
fewer than seven counts had optical counterparts.  Table~1 lists the 
IAU-approved name, position, X-ray counts, associated $1\sigma$ statistical 
uncertainty in the measured counts, and corrected count rate (counts s$^{-1}$) 
for the 99 {\it Chandra} sources in the $17\arcmin\times17\arcmin$ FOV. 
                        
The resulting 90\%-power ellipses were used to extract 99 individual 
event lists from which we extracted the hardness ratio, Kolmogorov-Smirnov 
time-variability statistic, light curves and spectrum for each source. 
We could only use second-epoch data for HHJ~257 (CXOP~J034655.6+235622) 
and first-epoch data for SCG94~171 (CXOP~J034716.9+241233), 
because of slight differences in pointing position. 
Additionally, it should be noted that HII~956 (CXOP~J034615.8+241122) dithered 
on and off the CCD during the second-epoch observation. 

\section{Data Analysis} 

\subsection{Source Matching} 

We constructed a database of all known 
optical \citep*[][USNO-A\footnote{United States Naval Observatory Catalog A 
2001 (USNO-A), available at 
http://www.fofs.navy.mil.}]{Mar00,Sta98,Ham93,Har82,Her47}, 
infrared \citep[][2MASS\footnote{2MASS 2001, 2d Incremental Release, Point 
Source Catalog, available at 
http://www.ipac.caltech.edu/2mass/releases/second/index.html.} 
]{Kri01,Mar00,Pin00,Sta98,Ham93}, and 
X-ray sources \citep{Mic99,Mic96,Sta94,CH85} in the ACIS-I FOV.  
The position uncertainties used to cross-reference previously cataloged 
sources are listed in Table~2.  {\it Chandra} position uncertainties are 
defined by the PSFs described above.  

The FOV contains 376 objects, 57 of which are unidentified {\it Chandra} 
sources.  Of the 99 {\it Chandra} sources, seven are only associated with 
known infrared sources, thirteen with only optical sources, and 28 with 
sources detected both in the optical and infrared.  There are no brown dwarf 
candidates in the {\it Chandra} FOV.  Alternate identifications and spectral 
data \citep[from][]{Mic96,Mic99,Har82} are listed in Tables~3 and 4.  
{\it Chandra} detected all 34 {\it ROSAT} sources in the ACIS-I FOV, 26 of 
which exhibit flare-like activity.  In Table~3 we cross-reference each 
{\it Chandra} source with all corresponding entries from previous catalogs.  
In six cases, single sources from other X-ray catalogs were resolved into two 
or three sources with {\it Chandra}. 

\subsubsection{Pleiades Members} 

{\it Chandra} detected eighteen of the 23 bona fide Pleiades members and all 
five of the possible Pleiades members in the ACIS-I FOV 
\citep{Pin00,Bel98,Sta98,Mic96,Sch95,Ham93}.  In this paper, Pleiades 
membership is defined by stars with a proper motion probability (ppm) $>50\%$ 
consistent with the known proper motion of the Pleiades cluster.  Possible 
members are stars with similar photometric properties as Pleiades members but 
where ppm$<50\%$.  In Figure~1, the Pleiades members are outlined on the ACIS-I 
image.  X-ray emission can be associated with five (of seven) early-type 
Pleiades members (B6~IV, A7~V, A9~V, F3~V, F4~V), and thirteen (of sixteen) 
late-type stars (K-M) during these observations \citep{Mic99,Mic96,Har82}.  

A spectral analysis of seven X-ray luminous Pleiades members revealed that 
there is a systematic, linear dependence of flux on the HR (see \S3.2) of a 
given source.  The flux conversion factor (FCF) can then be expressed as, 
\begin{equation} 
   {\rm FCF} = 1.5\times10^{-11} \,\times\, {\rm HR} \,+\, 1.03\times10^{-11}. 
\end{equation} 
Using this FCF, the X-ray luminosities ($L_{\rm X}$) were determined for all 
detected Pleiades members (for 0.5--8.0~keV). 

We calculated $L_{\rm X}$ upper limits for Pleiades members not 
detected by {\it Chandra}.  Assuming a constant background, we used detection 
limits corresponding to the faintest source in a given off-axis radial band 
($d$); the faintest sources where $0\arcmin\leq d<5\arcmin$ have 7~cts, 
$5\arcmin\leq d<10\arcmin$ have 10~cts, and $d>10\arcmin$ have 24~cts.  We 
then used the respective 95\% confidence level (where $2\sigma$ corresponds to 
a 97.72\% confidence level) upper limits for sources of $6_{-3}^{+6}$, 
$9_{-4}^{+7}$, and $23_{-7}^{+10}$ counts given by \citet{Geh86} to 
compute the $L_{\rm X}$ upper limits.  For these calculations we used a 
constant HR corresponding to the mean HR of detected Pleiades members 
(${\rm HR}=0.04$) and the effective exposure time at that source's location 
(see \S2.1). 

All of the undetected Pleiades members (HCG~254, HII~1362, HII~1375, MHO~10, 
and MHO~11) were located near the edge of the ACIS-I FOV.  The locations of 
undetected Pleiades members is illustrated by boxes in Figure~1.  As mentioned 
above ($\S2.2$), the point-source sensitivity of ACIS-I decreases as the PSF 
increases and effective area decreases with distance from the {\it Chandra} I3 
aim-point.  These effects may explain the non-detection of these Pleiades 
members.  It should be noted, however, that despite repeated observations 
\citep{Mic99,Mic96,Sta94,CH85}, these sources have not been detected with other 
X-ray observatories.  Thus these undetected sources, which must have low 
X-ray luminosities, might have been detected by {\it Chandra} if 
they were near the aim-point.  Table~5 shows spectral data and spectral types 
for all Pleiades members and possible members not detected during these 
{\it Chandra} observations.  

\subsection{Variability, Hardness Ratios and Spectra} 

The cleaned event files (\S2.1) list the arrival time, energy, and position of 
each X-ray photon.  A cumulative count distribution, $f_n(t)$, for each source 
was used to calculate the Kolmogorov-Smirnov statistic (KS) assuming a constant 
count rate, $f_o(t)$, \citep{BF96}, 
\begin{equation} 
   {\rm KS} = \sqrt{n} \,  \max |f_n(t)-f_o(t)|, 
\end{equation} 
where $n$ is the number of photon events and $t$ is the time of arrival. 
We computed the KS statistic for each source during the first and second 
observations to look for short term variability.  We used the CIAO tool 
{\em lightcurve} to extract light-curves for each source and found that 
short-term and flare-like variability was correlated with ${\rm KS} \geq1.00$. 
In Table~1, KS$_{s}$ is the larger of the KS values from 1999 September and 
2000 March. 

The $\sim6$ month interval between 1999 September and 2000 March allowed us to 
search for long-term variability.  We concatenated the two event lists by 
subtracting the $\sim6$ month gap from the second set of photon arrival times. 
The resulting long-baseline statistic, KS$_{l}$, is listed in Table~1.  We 
assume significant long-term variability and no significant short-term 
variability if KS$_{l} > 1.00$ and KS$_{s} < 1.00$.  There are 17 such sources 
in our sample.  As noted in $\S2.2$, KS$_{l}$ is not calculable for HII~956, 
HHJ~257, and SCG94~171 because the sources were only visible during one 
{\it Chandra} observation. 

We have compared the $L_{\rm X}$ values inferred for the six K stars and two F 
stars in our sample that were also observed with the ROSAT PSPC and analyzed by 
\citet{Gag95}. Five of the stars have nearly identical X-ray luminosities, but 
three of the K stars are factors of 2.1--2.5 times fainter than observed by 
ROSAT.  This magnitude and frequency of variations is consistent with the 
result found by \citet{Sch93} that $\sim 40$\% of the Pleiades sources are 
variable by a factor of 2 in an 11 year time span. 

The hardness ratio is defined by, 
\begin{equation} 
   {\rm HR = (hard-soft)/(hard+soft)}, 
\end{equation} 
where soft and hard counts are defined as ACIS events with energies between 
0.5-1.0~keV and 1.0-8~keV, respectively.  Soft sources show HR$<0.0$ and hard 
sources show HR$\geq0.0$.  We compute the error ($\sigma_{\rm HR}$) assuming
$1\sigma$ 
Poisson statistics from the combined event lists.  Table~1 tabulates KS$_{s}$, 
KS$_{l}$, HR, and $\sigma_{\rm HR}$ for each source. 

Using the variable-abundance APEC model in XSPEC version 11.0 we modeled ACIS 
spectra of the six most X-ray luminous Pleiades members in our FOV: 
HII~956 (A7 + F6), HII~980 (B6 + G), HII~1355 (K6~V), HII~1122 (F4 + K), 
HII~1124 (K3~V) and HII~1094 (K V).   The spectral differences between 
the 1999 and 2000 observations were significant enough that models of the 
combined data were not quantitatively accurate.  Therefore, the best-fit VAPEC 
parameters for each of the above stars, with $1\sigma$ uncertainties, listed 
in Table~6 correspond to the individual observations.  However, for a 
qualitative analysis, the ACIS spectra and model fits from the combined dataset 
are plotted in Figures~2a-f. 

These spectra were best fit with a column density of 
${\rm N_H=3.6\times10^{20}cm^{-2}}$, in agreement with the reddening 
to the cluster and previous column density estimates \citep{CH85,Sta94,Kri01}.  
For three stars (HII~1094, HII~1124, and HII 1355) single-temperature models 
yielded $\chi^2 > 2.0$.  Since the S/N of these spectra were not sufficient to 
constrain a two-temperature model with 9 free parameters, the Mg abundance was 
fixed at 1.00, the column density was fixed at 
$N_{\rm H}=3.6\times10^{20}{\rm cm}^{-2}$, and the upper temperature was fixed 
at $kT = 3.5$~keV.  We note that the excess emission near 0.7~keV in the 
spectrum of HII~956 (Fig.~2d) gave rise to $\chi^2 \approx 1.6$ for both 
one- and two-temperature models. 

\section{Results and Discussion} 

\subsection{Background AGN} 

Background active galactic nuclei (AGN) are generally detected in ACIS images, 
and the deep imaging capability of {\it Chandra} above 2--3~keV makes it 
possible to observe substantial populations of background AGN even through 
high column densities.  

Unfortunately, because the distant sources show few counts ($\sim7-171$ 
photons), we cannot reliably discriminate between AGN and faint background 
stellar populations from the shape of their spectra.  However, measurements of 
the HR and count rate are powerful tools for identifying background AGN, as 
they appear as hard sources with low count rates.  In Paper I 
\citep[and references therein]{Kri01}, we estimate that ultra-soft AGN would 
have ${\rm HR}>-0.32$, while most X-ray--bright AGN have ${\rm HR}>\sim0$. 
Stellar sources have ${\rm HR}\geq0.35$.  While HR alone cannot distinguish 
between stars and AGN, AGN typically have higher HR values.  Figure~3 
illustrates this population of unidentified, hard sources with low count rates. 

In the {\it Einstein} Medium-Sensitivity Survey (EMSS) \citep{Sto91}, AGN show 
$-1.0 \leq \log{(f_{x}/f_{v})} \leq 1.2$, whereas nearly all stars in the EMSS 
are much less active at X-ray wavelengths: $\log{(f_{x}/f_{v})} \leq -1.0$. 
To calculate the maximum expected X-ray flux from a star below the optical 
completeness threshold, we used the current upper limits \citep[V$=22.5$]{Sta98} 
and Equation~A3 from \citet{Mac88}, 
\begin{equation} 
   \log{(f_{x}/f_{v})}=\log{(f_{x})}+\frac{\rm V}{2.5}+5.37. 
\end{equation} 
A very X-ray active star at the completeness threshold of \citet{Sta98} 
would produce approximately three ACIS counts in our 56.4~ks {\it Chandra} 
observation.  We conclude that most of the unidentified X-ray sources are not 
coronal stars.  
Table~4 also shows a population of ten optically faint (${\rm V}>22.5$, 
${\rm R}>20.2$ or ${\rm J}>14.5$) X-ray sources whose $\log{(f_{x}/f_{v})}$ 
values also exceed -1.  Figure~3 illustrates that both the unidentified X-ray 
sources and most of the optically faint X-ray sources have similar X-ray 
properties (low count rates, low KS values, and high HR values).  At this flux 
level and at this galactic latitude, the most likely candidates for both 
populations are the optically faint, X-ray--bright AGN. 

We also note that the candidate AGN do not show medium- to large-amplitude 
short-term variability, as seen on the late-type stars, and have low KS values. 
This is consistent with known AGN activity since AGN generally show only 
low-amplitude variability on time-scales $\leq1$~day \citep{Zam84}.  For 
faint AGN, this type of low-level variability would not be evident in a light 
curve and would yield low KS statistics. 

A {\it Chandra} observation of the Hubble Deep Field-North \citep{Hor01} 
analyzed the X-ray properties of 82 sources associated with the 
$8.6\arcmin \times 8.7\arcmin$ area covered by the Caltech Faint Field Galaxy 
Redshift Survey.  We compared our data to those of \citet{Hor01} by assuming 
${\rm FCF_{AGN}=1.33\times10^{-11}\,ergs\,cm^{-2}\,s^{-1}\,per\,ct\,s^{-1}}$ 
for a typical AGN candidate.  We found that the AGN candidates in our 
observation have full band ($0.5-8.0$~keV) fluxes of 
$4\times10^{-14} < {\rm f_{AGN}} < 2\times10^{-15}$~ergs~cm$^{-2}$~s$^{-1}$. 
These flux levels encompass 26 sources in $\sim\frac{1}{4}$ of the ACIS-I FOV 
\citep{Hor01}.  Therefore, in the direction of the Pleiades, we expect to find 
$<100$ AGN in our FOV.  The column density through the Galaxy, in the direction 
of the Pleiades is N$_{\rm H} \approx 1.1\times10^{21}$cm$^{-2}$ \citep{Dic90}. 

Our source detections give a population of fifty-seven AGN candidates that have 
not been detected at any other wavelength, and ten optically 
faint AGN candidates; all of these show HR$\geq-0.32$ and KS$_{s}\leq1.5$.  
Five of these have previously been detected with {\it ROSAT} \citep{Sta94}.  
Two of the AGN candidates show HR$\geq0.35$.  In Figure~3 it is obvious that 
stellar sources (namely Pleiades members) populate the same region of the 
diagram as candidate AGN.  The reader should note the relatively large 
$\sigma_{\rm HR}$ values associated with these sources.  Every source in 
Table~1 without an optical or infrared indentification is an AGN candidate.  
Additionally, X-ray detected sources with measured optical and/or infrared 
magnitudes and with $-1.0\leq\log{(f_{x}/f_{v})}\leq1.2$ are marked as AGN 
in Table~4. 

\subsection{Are B- and A-type stars intrinsic X-ray sources?} 

Of the 23 bona fide Pleiades members present in the ACIS-I FOV, {\it Chandra} 
detected eighteen including: B6~IV + G, A7~V + F6, A9~V + K, F3~V + F6, 
F4~V + K, seven additional K-type, and five M-type stars.  One detected but 
optically 
faint Pleiades member (HHJ~92) had no listed spectral type, but its color 
indicates that it is a late-type star.  Based on this sample of stars, we 
examine the question of whether B4-F5 type stars are intrinsic X-ray emitters, 
or if the X-ray emission observed from some of these stars comes from late-type 
companions.  

Theoretical stellar structure models \citep[e.g.,]{Boh84} show that 
sub-photospheric convective zones decrease in thickness with increasing 
$T_{\rm eff}$ and disappear entirely for main sequence stars at $T_{\rm eff} 
\approx 9000$~K, corresponding to spectral type A2~V.  Although the precise 
temperature at which convection disappears depends on gravity, metal abundance, 
the mixing length, or other convection parameters, all theories show a rapid 
decrease in convective zone thickness from the F stars to the early-A stars. 
This is important because the presence of a turbulent convective layer is 
usually thought necessary to drive a magnetic dynamo.  The turbulent motions in 
a convective zone also continually shuffle the foot-points of magnetic flux 
tubes, leading to field reconnection events (flares and micro-flares) in the 
corona and thus heating.  In an atmosphere, these convective motions propagate 
outward, leading to heating by the dissipation of acoustic and MHD waves.  
Field reconnection events and wave dissipation are the mechanisms typically 
assumed to heat the chromosphere and coronae of low mass stars.  

However, in 1991, \citet{SL91} used the IUE to find chromospheric 
C~{\scriptsize II}~1335{\AA} emission in late-A to early-F stars.  Soon after, 
\citet*{WML95} found C~{\scriptsize II}~1335{\AA} from $\alpha$~Aql (A7~IV-V) 
and $\alpha$~Cep (A7~IV-V).  The discovery of N~{\scriptsize V}~1239{\AA} 
emission from $\alpha$~Aql and $\alpha$~Cep, and Si~{\scriptsize III}~1206{\AA} 
emission from $\alpha$~Aql, $\alpha$~Cep, and $\tau^{3}$~Eri (A4~V) 
\citep{SL97} confirmed that some stars as early as A4~V have an active upper 
atmosphere with plasma heated to between 30,000~K and 150,000~K. 
Additional observations by \citet{Rac97,Rac98,Rac00} confirmed chromospheric 
He~{\scriptsize I}~5876{\AA} (D3) emission in A5--F5 (III--V) stars in the 
Hyades, Praesepe, Coma, the Pleiades and Alpha Persei.  In sum, there exist 
some late-A to early-F type stars with confirmed chromospheres.  This, coupled 
with the findings of \citet[][\S1]{Wal83}, implies that either (1) the shallow 
convective zones of these stars are able to heat the upper atmospheres through 
the dissipation of acoustic and/or MHD waves, or (2) there exists some heating 
mechanism that is not completely understood or fully incorporated into current 
theoretical models. 

Massive stars are not expected to show dynamo-driven magnetic 
activity, but the spectral type at which the magnetic or wave mechanisms for 
coronal heating disappear is not yet identified with any precision.  Strong 
wind shocks on O- and early B-type stars lead to relatively soft, non-variable 
X-ray emission in high-mass stars \citep*{OC99, Owo88, LW80}.  Therefore, 
current models do not predict X-ray emission from late-B to A-type stars. 
However, Paper I, \citet{Hue01}, \citet{Sta94}, \citet{Cai94}, 
\citet{BS94}, \citet{Sch93}, \citet{Mic90}, \citet*{Pal90}, \citet{CZ89}, and 
\citet{Sch85} have all recorded X-ray emission from a small percentage of the 
observed B and A stars using the {\it Einstein}, {\it ROSAT}, and {\it Chandra} 
satellites.  \citet{Mic96}, \citet{Sta94}, \citet{Sch93}, \citet{Gri92}, 
\citet{CH85}, and \citet{Gol83} have proposed that this X-ray emission from 
some early-type stars could be explained as emission from previously unknown 
late-type companions. 

The hypothesis that late-B and A-type X-ray sources have late-type coronal 
companions is difficult to refute on the basis of X-ray luminosity or detection 
fraction alone.  \citet{Sta94} noted that the B- and A-type star X-ray 
luminosities are comparable to active G- and K-type stars, and the fraction of 
X-ray emitting B- and A-type stars does not exceed the fraction of B- and 
A-type stars expected to have close late-type companions.  However, with the 
increased S/N and wider bandpass of {\it Chandra}, we can revisit the question 
of B- and A-type star X-ray emission by comparing the X-ray characteristics of 
B- and A-type X-ray sources with flaring and non-flaring Pleiades 
members.  Since the Pleiades stars are likely co-eval, late-type 
companions to early-type stars should have the same X-ray properties as single 
late-type stars.  A comparison of the hardness ratios, KS statistics, 
light curves, X-ray luminosities, and X-ray spectra of early- and late-type 
Pleiades stars thus should test the hypothesis.  In order to do this, we split 
the Pleiades sources into three categories with similar X-ray properties:  
low X-ray luminosity stars, high X-ray luminosity K stars, and high X-ray 
luminosity late-B to early-F stars.  X-ray characteristics for these groups are 
summarized in Table~7. 

\subsubsection{Low X-ray Luminosity Stars} 

The X-ray faint (inactive) Pleiades members in Table~7 are characterized by 
$\log{L_{\rm X}} < 28.2$ and ${\rm HR} \approx0.0\pm0.2$.  HII~1284's 
(CXOP~J034704.0+235942) (A9~V + K) X-ray properties ($\log{L_{\rm X}}=27.58$ 
and ${\rm HR} = -0.20\pm0.16$) are consistent with X-ray emission from an 
inactive late-type companion.  HII~1284 has been repeatedly observed 
\citep{Mic99,Mic96,Sta94,CH85} and not detected in X-rays. 

\subsubsection{Active K-type Stars} 

All X-ray luminous K-type stars in our sample, except for HII~1124, were 
observed to flare, as shown by Table~7.  These stars also have hardness ratios 
with values between $-0.28\leq{\rm HR}\leq-0.15$.  Two-temperature VAPEC models 
\citep{Smi01} are needed to adequately fit the X-ray spectra of most of the 
active K-type stars (see Figures~2a-c, Table~6, and \S3.2), HII~1094 (K~V), 
HII~1124 (K3~V), and HII~1355 (K6), the exceptions being observations during 
which there were no large-scale flares.  Presumably, these plasma components 
correspond to cooler coronal emission at ${\rm kT\approx0.3-0.6~keV}$ and 
hotter ${\rm kT\approx3.5~keV}$ plasma associated with high- or low-level 
flaring.  Model fits to the HII~1094, HII~1124, and HII~1355 X-ray data
also revealed sub-solar iron abundances relative to solar 
($0.17$--$0.66$, $0.13$--$0.30$, and $0.04$--$0.70$ 
respectively) \citep{And89}.  Both HII~1094 and 
HII~1355 show super-solar neon ($2.05$--$2.95$ and $1.22$--$2.91$) 
and silicon ($1.5$--$2.5$ and $2$--$4$) abundances (see Table~6).  The reader 
should note that the spectral resolution of ACIS-I is relatively low and, 
therefore, the derived abundances should be interpreted with caution.  However, 
similar abundance patterns are seen in high-resolution grating spectra of the 
Pleiades moving-group member AB~Doradus (K1~V) and in other active late-type 
stars \citep{Gud01,Dra01}.  

\subsubsection{HII~956 (A7~SB? + F6)} 

HII~956 is a non-flaring source, has a high X-ray luminosity 
($\log{L_{\rm X}}=29.31$), consistent with previous measurements 
\citep{Mic96,Mic90,Sta94,CH85}, and a soft spectrum (HR$=-0.48\pm0.03$).  As 
noted earlier, we find a marginally acceptable one-temperature fit with 
$kT=0.57~{\rm keV}$ (see Figure~2d and Table~6).  This spectrum also requires 
somewhat enhanced abundances (where ${\rm Ne=1.5\pm0.3}$ and 
${\rm Mg=1.5\pm0.3}$ relative to solar). 

The A7~V primary of HII~956 is a suspected spectroscopic binary \citep{Liu91}. 
It has a visual companion in the Tycho catalogue with separtion 
$0.85\arcsec$, V=9.45 and B$-$V=0.54. The visual companion (at 110 A.U.) is 
very similar to the F6 Pleiad HII~1309 (K. Briggs \& J. Pye, private 
communication). We conclude that the X-rays are likely emitted by the 
late-F-type visual companion. 

\subsubsection{HII~980=Merope (B6 + G?)} 

HII~980 is the most bolometrically luminous star in the {\it Chandra} 
FOV, with $\log{L_{\rm X}}=29.60$.  The X-ray spectrum of HII~980 
(Figure~2e and Table~6) requires one plasma with temperatures ranging from 
${kT=0.51}$--${\rm 0.61~keV}$.  HII~980 suggests sub-solar iron 
($0.58$--$0.86$), super-solar neon ($1.0$--$1.5$), magnesium ($1.30$--$1.89$), 
and silicon ($1.30$--$2.41$) abundances.  We note, however, that 
VAPEC fits with solar Ne, Mg, and Si abundances yield only slightly poorer 
$\chi^{2}$ values.

\citet{Dom00} report a {\it Hipparcos} visual companion to Merope, with 
no further information. Merope's X-ray properties are consistent with a 
moderately active early-G-type companion \citep{Sta94,Gud97}. The {\it Chandra} 
light curve shows some short-term and long-term variability, but no obvious 
flares. Historically \citep{Mic96,Mic90,Sta94,CH85}, HII~980 has displayed 
steady, bright X-ray emission ($\log{L_{\rm X}}\approx29.8$). 

\subsubsection{HII~1122 (F4 + K) and HII~1338 (F3 + F6)} 

HII~1122 (F4~V + K) and HII~1338 (F3~V + F6) have X-ray luminosities 
consistent with those reported by \citet{Sta94, Mic99, Mic96, Mic90}. 
We measure $L_{\rm X}=29.06$ for HII~1122 and $L_{\rm X}=28.67$ for HII~1338.  
These stars have hardness ratios (${\rm HR}\approx-0.45$) comparable to 
those of HII~956 (${\rm HR}=-0.48\pm0.03$) and HII~980 
(${\rm HR}=-0.34\pm0.02$).  The light curves of both HII~1338 and HII~1122 are 
similar to those of HII~956 and HII~980 as shown in Figure~4.  The spectrum of 
HII~1122 (Figure~2f and Table~6) indicates a temperature of 
$kT=0.45~{\rm keV}$, like HII~956 and HII~980.  Note also the super-solar 
magnesium (${\rm 2.79\pm0.74}$) abundance. 

HII~1122 is a double-lined spectroscopic binary with a secondary mass 
$M \approx0.5M_{\odot}$ \citep{Liu91,Rab98}, suggesting a K-type secondary. 
K stars with $L_{\rm X}\approx29$ generally show hard X-ray spectra, higher 
hardness ratios, and occasional flaring. HII~1122 is the softest X-ray 
source in our survey. We suggest that the F4 primary is the source of the X-ray 
emission. 

HII~1338 is a double-lined spectroscopic binary \citep{Liu91} with well 
determined orbital elements \citep{Rab98}. The spectral type and luminosity 
of the primary imply a highly inclined orbit and a secondary mass of 
$M\approx1.3M_{\odot}$ (K. Briggs \& J. Pye, private communication). 
Based on its hardness ratio and $L_{\rm X}$, the F6 secondary is the most 
likely X-ray source; although the F3 star may contribute to the X-ray emission. 
Hence, mid-F stars appear to be strong, soft X-ray sources as evidenced by 
HII~956B (F6), HII~1122A (F4), and HII~1338B (F6). 

Among the moderately active F, G, and K stars listed in Table~7, 
we see a clear increase in hardness ratio with decreasing mass of the 
X-ray emitter. Active M-type stars on the other hand 
appear to have somewhat cooler coronae than their active K-type counterparts. 
Low-resolution PSPC spectra of the Pleiades suggest a similar result: 
the hardest X-ray sources are the rapidly rotating, flaring G stars, 
followed by K stars, slowly rotating G stars, and F stars \citep{Gag95}. 

\subsubsection{Undetected A stars: HII~1375 (A0 + A) and HII~1362 (A7)} 

{\it Chandra} did not detect HII~1375 (A0~V + A~SB) or HII~1362 (A7) with a 
very low upper limit of $\log{L_{\rm X}}<27.98$ and $\log{L_{\rm X}}<27.87$ 
respectively.  As these stars have been repeatedly observed and not detected 
in the X-rays \citep[$\S3.1.1$]{Mic99,Mic96,Mic90,Sta94,CH85}, we conclude that 
they are not strong X-ray sources and that they do not have active late-type 
companions.  It is interesting to note that while HII~1362 is one of the 
earliest known stars to have a detectable chromosphere \citep{Rac00}, it was 
not detected by {\it Chandra}. 

\subsubsection{Detection fraction of flares} 

As reported by \citet{Gag95}, flares detected by {\it ROSAT} exhibited peak 
X-ray luminosities in excess of $L_{\rm X}=3\times10^{30}\,{\rm ergs\,s^{-1}}$. 
Only ten flares were detected from 171 Pleiades members \citep{Gag95,Sta94} 
with {\it ROSAT} (all associated with late-type stars).  As a result of its 
small PSF, increased effective area, and low ACIS-I background, {\it Chandra} 
detected flare-like activity on nine of twenty-three Pleiades members on 
similar time-scales. 

{\it Chandra} reveals flare-like activity on five of the seven K-type 
Pleiades members in the FOV.  Also, of the seven early-type Pleiades members 
in the FOV, {\it Chandra} detected five; none of which show flare-like 
variability.  In the cases of HII 1284 (A9 + K) and HII~980 (B6 + G), 
the late-type secondaries are relatively inactive and less likely to flare. 
In the cases of HII~956B (F6), HII~1122A (F4), and HII~1338B (F6), the lack 
of short-term variability is consistent with previous observations of F stars 
with the PSPC \citep{Gag95}. 

\subsection{Possible New Pleiades Members} 

{\it Chandra} detected all five of the possible Pleiades members in the 
ACIS-I FOV.  Membership probability was determined by proper motion and 
photometric properties in agreement with the Pleiades cluster 
\citep[$\S3.1$]{Bel98,Sta98,Mic96,Sch95,Ham93,Sod93}.  Tentative X-ray 
luminosities have been determined for the possible Pleiades members using the 
same parameters (including distance) as for the bona fide Pleiades members.  
These X-ray characteristics are listed in Table~7. 

\subsubsection{HHJ~140 (M~V)} 

The light-curve of HHJ~140 (CXOP~J034635.4+240134) (M~V) displays a high 
amplitude flare during the second observation, exceeding the count rate of the 
first observation by a factor of 75.  This hard source shows HR$=0.17\pm0.03$ 
and $\log{L_{\rm X}}=29.48$.  This is much harder than typical HR values of the 
active Pleiades late-type stars.  However, the hardness of HHJ~140's spectrum 
is due to the flare: during the first observation HR$=-0.59$, and 
during the second HR$=0.22$.  As the second observation contains the majority 
of the counts, the overall HR value is heavily weighted by the second 
observation.  Since no other sources in these observations show flares of this 
magnitude, we cannot directly compare the X-ray properties of this source with 
Pleiades stars.  However, the large flare and X-ray luminosity indicate that 
HHJ~140 is very active, and thus is consistent with being a Pleiades member. 

\subsubsection{HHJ~195 and HHJ~257} 

HHJ~195 (CXOP~J034623.3+240151) and HHJ~257 (CXOP~J034655.6+235622) are
both faint X-ray sources showing $\log{L_{\rm X}}=28.13$ and 
$\log{L_{\rm X}}=28.32$, respectively, if they are located at the distance of 
the Pleiades.  The HR values (HR$=-0.23\pm0.11$ and HR$=-0.23\pm0.16$ 
respectively) are also within the range for X-ray faint late-type Pleiades 
members.  These stars have been speculatively included in two membership 
catalogs \citep{Sta94,Ham93}, but do not have associated membership 
probabilities.  The X-ray properties indicate that these stars lie within the 
range observed in the Pleiades cluster and are thus likely Pleiades members.  

\subsubsection{SRS~62618 and SRS~60765} 

SRS~62618 (CXOP~J034643.5+235941) and SRS~60765 (CXOP~J034709.1+240307) are
candidate Pleiades members in the \citet{Sch95} and \citet{Bel98} catalogs.  
These sources have low proper motion probability, although their photometric 
properties are consistent with Pleiades membership (see Table~4 and $\S$3.1.1). 
SRS~62618 and SRS~60765 are relatively soft sources (HR$=-0.23\pm0.07$ and 
HR$=-0.19\pm0.16$), with high X-ray luminosities ($\log{L_{\rm X}}=28.67$ and 
$\log{L_{\rm X}}=29.03$), and high KS$_{s}$ values (KS$_{s}=1.33$ and 
KS$_{s}=3.50$).  These stars have X-ray properties of active late-type Pleiades 
members, suggesting that SRS~62618 and SRS~60765 are Pleiades members and not 
field stars. 

\section{Summary} 

Based on our analysis of the 1999 September and 2000 March {\it Chandra} ACIS-I 
observations of the core of the Pleiades cluster, 

\begin{enumerate} 

\item{We identify 99 X-ray sources, of which 57 have not been 
detected at any other wavelength.  Hardness ratios, Kolmogorov-Smirnov 
statistics, and count rates have been derived for all 99 {\it Chandra} sources.}

\item{The 57 sources are faint, hard, and have low KS values and are not listed 
in optical/IR catalogs.  They also show $-1.0 \leq \log{(f_{x}/f_{v})} 
\leq 1.2$. There are an additional ten faint optical/IR sources with similar 
$\log{(f_{x}/f_{v})}$.  These 67 sources are probably background AGN.} 

\item{Four {\it Chandra} sources are mid-B to early-F type stars with cooler 
companions: HII~980 (B6+G), HII~956 (A7+F6), HII~1338 (F3+F6), and HII~1122 
(F4+K). These sources exhibit consistently high X-ray luminosity, soft X-ray 
spectra, and little or no detectable short-term variability. All of these 
sources are spectroscopic or visual binaries. The B6 and A7 primaries in 
HII~980 and HII~956 are probably not intrinsic X-ray sources. The X-ray 
emitters are probably all F and G-type stars --- HII~980B, HII~956B, HII~1338B, 
and HII~1122A. This conclusion is strengthened by the nondetection of A stars 
without cooler companions, HII~1375 (A0V + A~SB) and HII~1362 (A7), which have 
X-ray luminosity upper limits 27--54 times smaller than the B6--A7 stars with 
cooler companions, HII~980 (B6 + G) and HII~956 (A7~SB? + F6).} 

\item{Among the sample of active F, G and K-type Pleiades members listed in 
Table~7, we see a clear anti-correlation between the mass of the X-ray emitter 
and its hardness ratio, although active M-type stars appear to have somewhat 
cooler coronae than their active K-type counterparts.} 

\item{HII~1284 (A9 + K) has X-ray properties comparable to those of K-type 
Pleiades members. Its X-ray emission probably comes from HII~1284B.} 

\item{Five Pleiades members are not detected in the combined Chandra image, 
putting stringent upper limits on the X-ray luminosity of some early- and 
late-type cluster members.   Two of these are Pleiades A stars.} 

\item{HHJ~140, SRS~62618, SRS~60765, HHJ~195, and HHJ~257 are likely Pleiades 
members.} 

\end{enumerate} 

\acknowledgments 

We would like to thank John Pye and Kevin Briggs for alerting us to the binary 
nature of the B-, A-, and F-type stars in the {\it Chandra} field of view. 
This study was made possible by NASA grant H-04630D to NIST and the University 
of Colorado.  We are grateful to Anita Krishnamurthi for providing compiled 
catalogs of sources in this region, and for her helpful comments on imaging.  
We thank Chris Reynolds for his helpful comments on AGN.  We would also 
like to thank Takamitsu Miyaji for providing the code to eliminate flaring 
pixels. This research made use of the SIMBAD astronomical database operated 
by CDS at Strasbourg, France.

\clearpage 
\begin{figure} 
\figurenum{1} 
\plotone{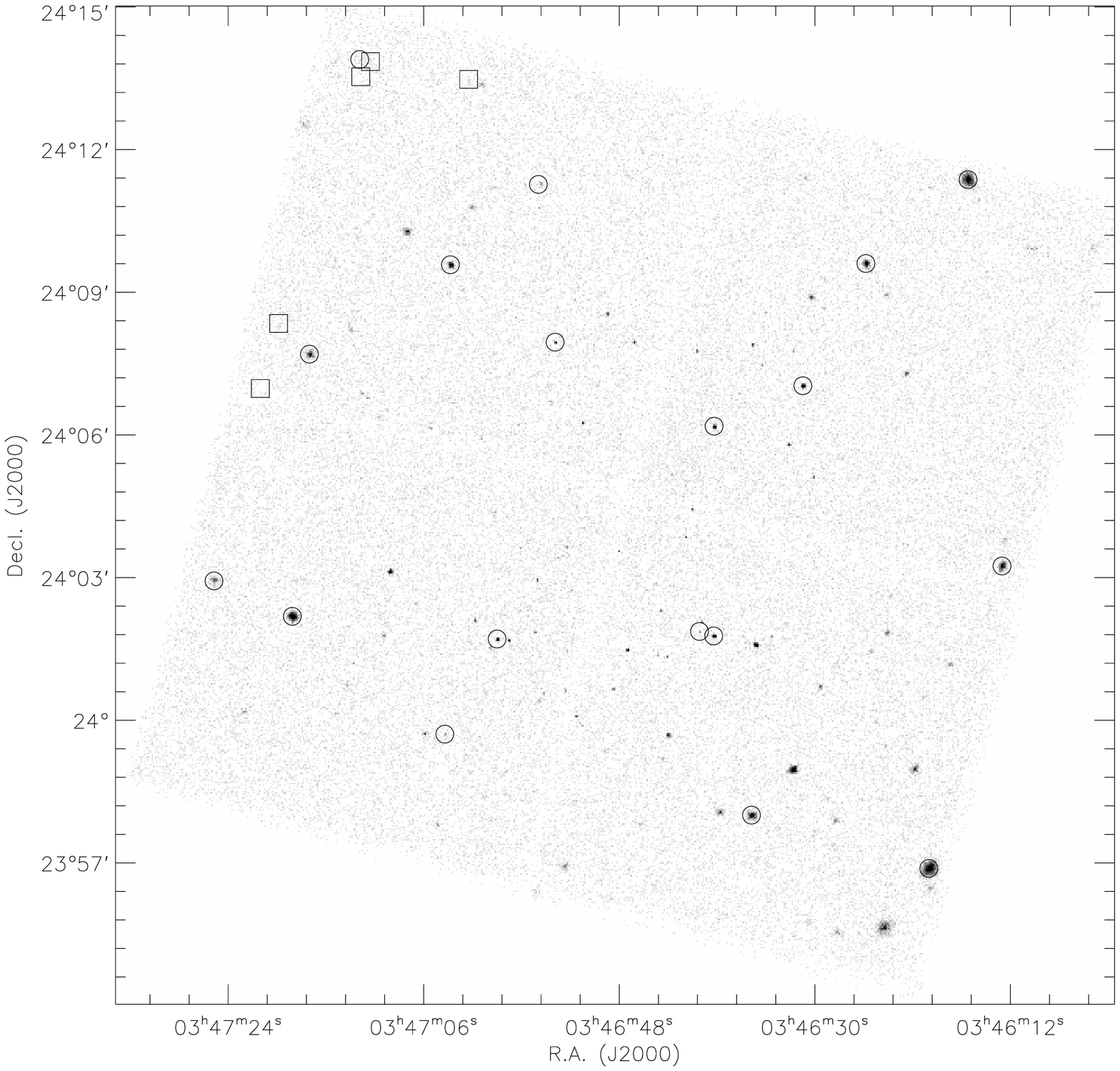} 
\caption{This co-added 56.4 ks image of the Pleiades core was taken with 
        ACIS-I on board {\it Chandra}.  Detected Pleiades members are enclosed 
        by circles and undetected Pleiades members are enclosed by squares.} 
\end{figure} 

\clearpage 
\begin{figure} 
\figurenum{2} 
\plotone{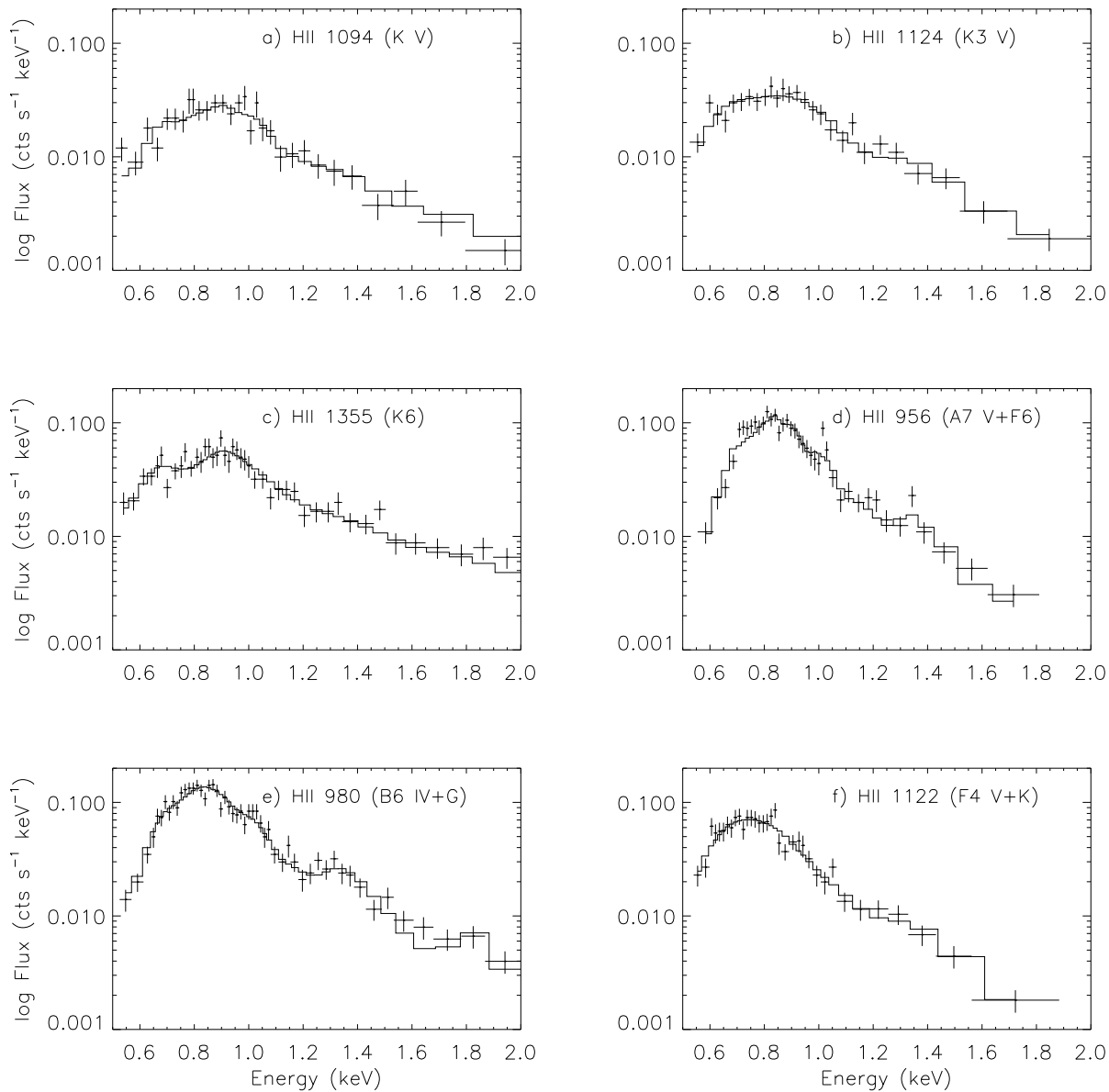} 
\caption{Spectra of the six most X-ray luminous Pleiades members in the ACIS-I 
        FOV, assuming a column density of ${\rm N_H=3.6\times10^{20}cm^{-2}}$ 
        to the Pleiades.  HII~980, HII~956, and HII~1122 required a single 
        temperature coronal model, while the K-type stars required two 
        temperature models.} 
\end{figure} 

\clearpage 
\begin{figure} 
\figurenum{3} 
\epsscale{.45} 
\plotone{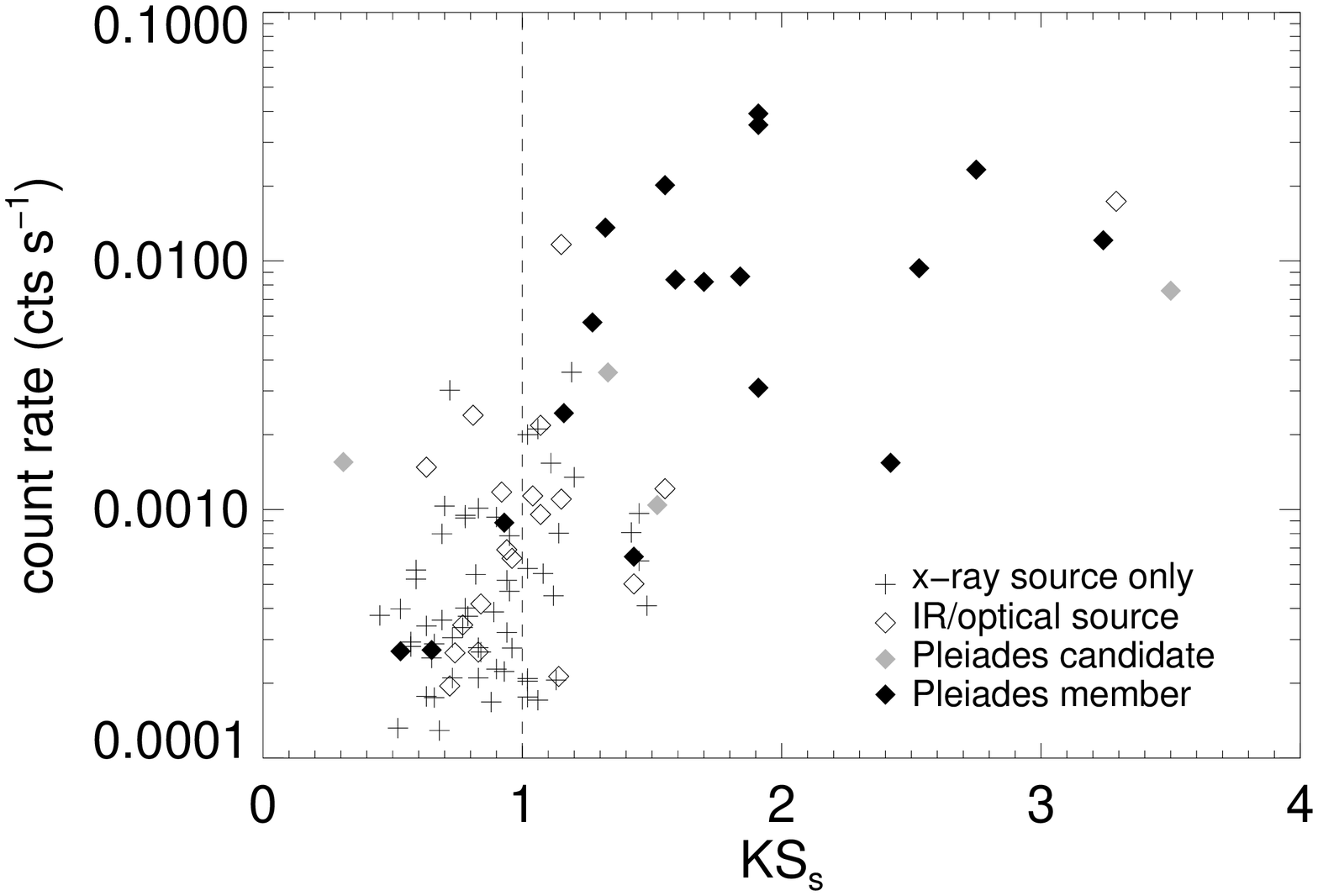} 
\plotone{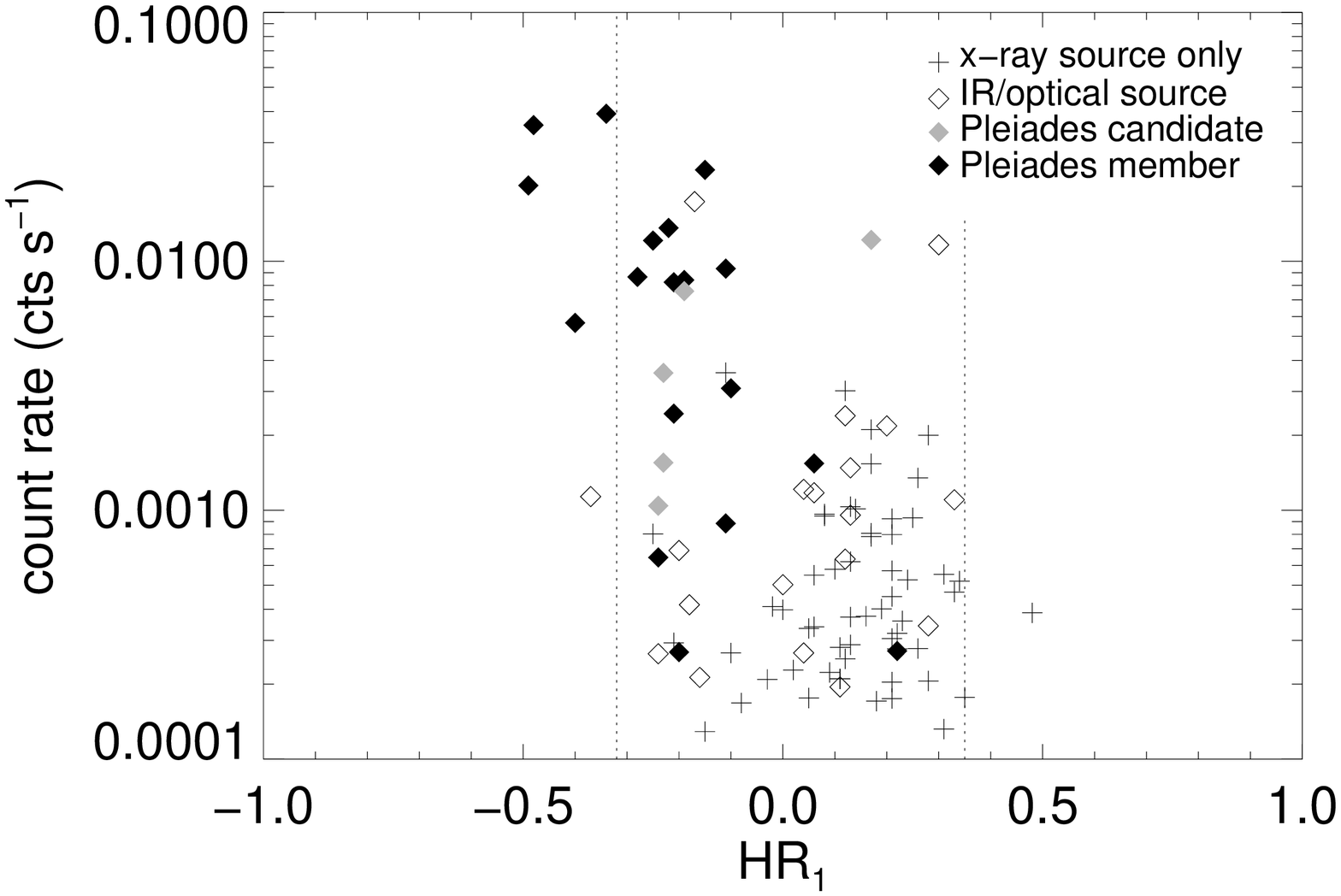} 
\plotone{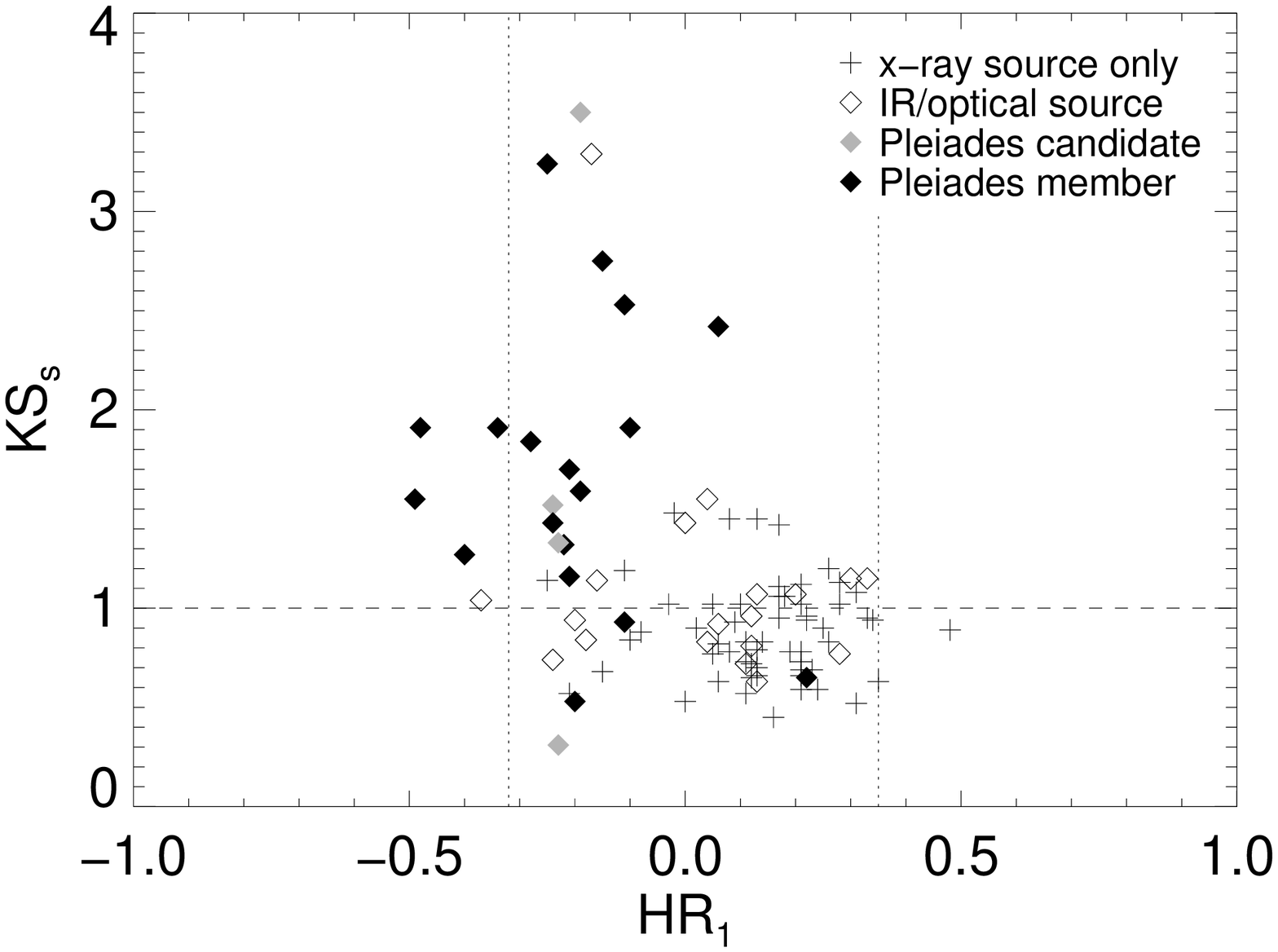} 
\caption{These plots of the KS statistic, count rates, and HR values are 
        useful tools in detecting AGN.  Ultra-soft AGN could have HR values 
        as small as $-0.32$, most X-ray--bright AGN will show HR values greater 
        than $\sim0$, and sources where HR$\geq0.35$ are almost 
        certainly AGN.  Dotted lines indicate these critical values.  The 
        dashed lines mark the threshold between variable and non-variable.  
        Note the population of hard, faint, non-variable sources.} 
\end{figure} 

\clearpage 
\begin{figure} 
\figurenum{4} 
\epsscale{1} 
\plotone{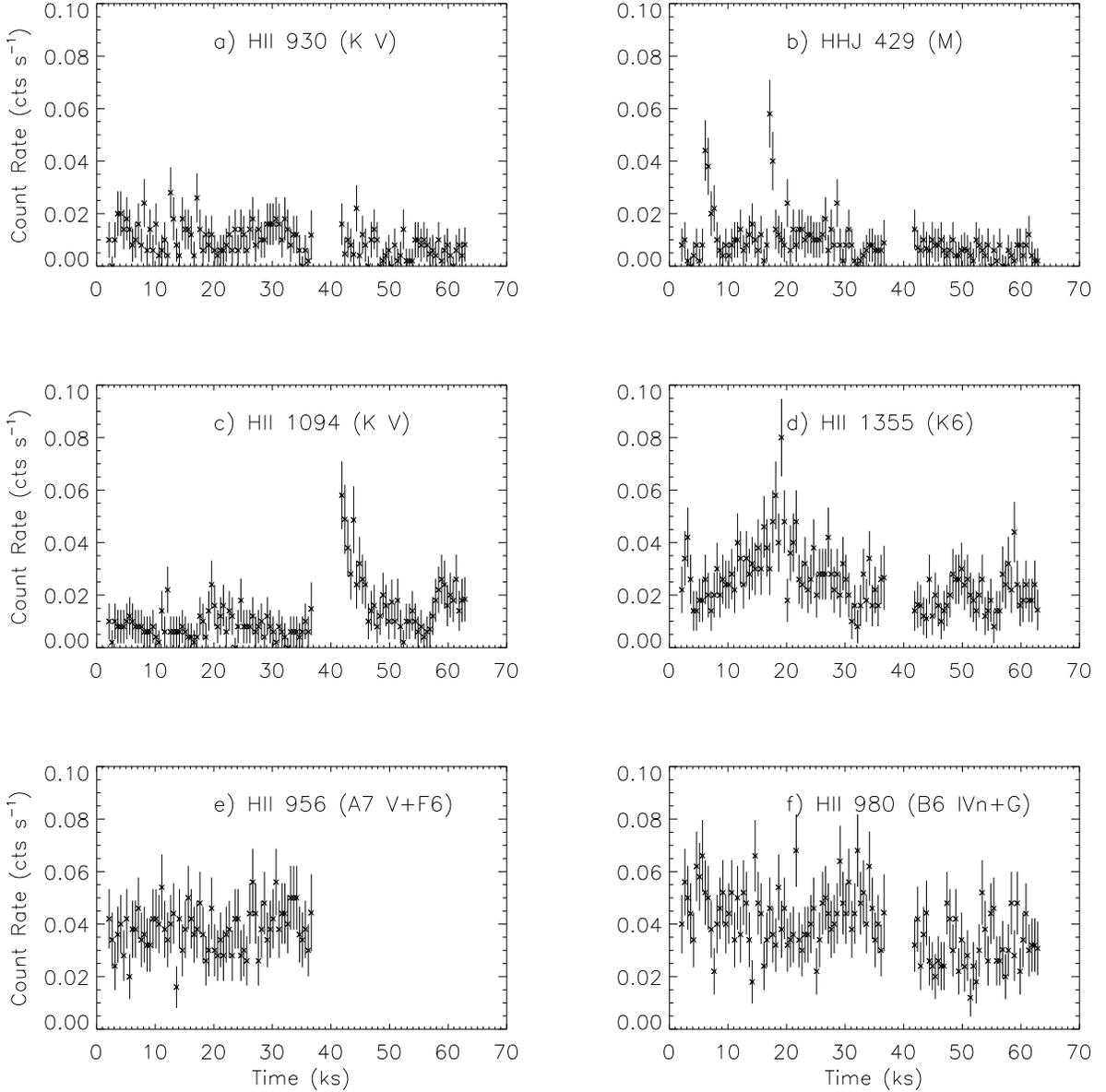} 
\caption{ACIS X-ray light-curves of active K-type stars and early-type 
        stars (500~s bins).  The data gap represents the six month interval 
        between observations.  All of the highly X-ray luminous 
        K stars show flare-like activity.  Light-curves (e) and (f) show 
        undulating, short-term variability over the course of a single 
        observation, and larger variability from 1999 September 18 to 2000 
	March 20.  HII~956 (e), HII~980 (f), HII~1122 (not shown), and HII~1338 
	(not shown) have not exhibited short-term flare-like variability in any 
	of the numerous observations with {\it Einstein}, {\it ROSAT}, and now 
        {\it Chandra}. } 
\end{figure} 

\clearpage 
\begin{deluxetable}{lrrrrrrrrrl} 
\rotate 
\tablecolumns{11} 
\tabletypesize{\scriptsize} 
\tablewidth{0pt} 
\tablecaption{{\it Chandra} X-ray sources in the core of the Pleiades.} 
\tablehead{ 
\colhead{CXOP}                 & \colhead{R.A.}               & 
\colhead{Decl.}                & \colhead{counts}             & 
\colhead{error}                & \colhead{count rate}         & 
\colhead{KS$_{s}$}             & \colhead{KS$_{l}$}           & 
\colhead{HR}                   & \colhead{$\sigma_{\rm HR}$}  & 
\colhead{Name\tablenotemark{a}}\\ 
\colhead{}                     & \colhead{(J2000)}            & 
\colhead{(J2000)}              & \colhead{}                   & 
\colhead{}                     & \colhead{(cts~s$^{-1}$)}        & 
\colhead{}                     & \colhead{}                   & 
\colhead{}                     & \colhead{}                   & 
\colhead{}                     } 
\startdata 
J034604.5+240957&  3 46 04.59& 24 09 57.9&    32&   10& 0.000689& 0.94& 0.49&-0.20& 0.10&BPL~88 \\ 
J034609.9+240541&  3 46 09.92& 24 05 41.7&    11&    8& 0.000228& 0.90& 1.79& 0.02& 0.10&\nodata \\ 
J034610.0+240954&  3 46 10.01& 24 09 54.6&    45&    9& 0.000948& 0.78& 0.63& 0.08& 0.07&\nodata \\ 
J034612.5+240347&  3 46 12.60& 24 03 47.3&    22&    6& 0.000450& 1.12& 1.59& 0.21& 0.10&\nodata \\ 
J034612.7+240314&  3 46 12.77& 24 03 14.6&   411&   22& 0.008411& 1.59& 3.71&-0.19& 0.04&HII~930 \\ 
J034615.8+241122&  3 46 15.90& 24 11 22.0&  1344&   39& 0.035244& 1.91& ...&-0.48& 0.03&HII~956 \\ 
J034617.6+240110&  3 46 17.61& 24 01 10.2&    38&    8& 0.000784& 0.95& 0.79& 0.17& 0.10&\nodata \\ 
J034619.4+235628&  3 46 19.45& 23 56 28.9&    24&    8& 0.000621& 1.45& 1.05& 0.13& 0.10&\nodata \\ 
J034619.5+235653&  3 46 19.51& 23 56 53.0&  1825&   46& 0.039150& 1.91& 4.73&-0.34& 0.02&HII~980 \\ 
J034620.1+240028&  3 46 20.10& 24 00 28.8&    13&    5& 0.000267& 0.84& 0.59&-0.10& 0.13&\nodata \\ 
J034620.8+235858&  3 46 20.85& 23 58 58.5&   171&   16& 0.003567& 1.19& 2.12&-0.11& 0.06&\nodata \\ 
J034621.5+240717&  3 46 21.58& 24 07 17.3&    68&   10& 0.001348& 1.20& 0.96& 0.26& 0.10&\nodata \\ 
J034623.3+240151&  3 46 23.37& 24 01 51.1&    52&    9& 0.001042& 1.52& 1.36&-0.24& 0.11&HHJ~195 \\ 
J034623.4+240856&  3 46 23.42& 24 08 56.9&    27&    6& 0.000552& 1.08& 0.77& 0.31& 0.13&\nodata \\ 
J034623.6+235538&  3 46 23.64& 23 55 38.9&   542&   26& 0.011645& 1.15& 1.35& 0.30& 0.04&0346237+235540 \\ 
J034624.3+235958&  3 46 24.38& 23 59 58.5&    13&    5& 0.000277& 0.96& 1.43& 0.22& 0.19&\nodata \\ 
J034624.8+240126&  3 46 24.88& 24 01 27.0&    16&    6& 0.000320& 0.94& 1.45& 0.22& 0.17&\nodata \\ 
J034625.2+240936&  3 46 25.30& 24 09 36.2&   448&   23& 0.009348& 2.53& 4.51&-0.11& 0.04&HHJ~429 \\ 
J034627.9+235532&  3 46 27.95& 23 55 32.1&    57&   10& 0.001213& 1.55& 0.92& 0.04& 0.08&0346379+235525 \\ 
J034628.1+235753&  3 46 28.13& 23 57 53.7&    50&    9& 0.001033& 0.70& 1.47& 0.13& 0.09&\nodata \\ 
J034628.8+235820&  3 46 28.89& 23 58 20.8&    13&    5& 0.000267& 0.83& 0.67& 0.04& 0.13&0346290+235822 \\ 
J034629.1+240839&  3 46 29.18& 24 08 39.9&    10&    4& 0.000195& 0.72& 0.54& 0.11& 0.24&0346290+240839 \\ 
J034629.5+240042&  3 46 29.53& 24 00 42.8&    51&    8& 0.001012& 0.83& 0.54& 0.14& 0.11&\nodata \\ 
J034630.0+235740&  3 46 30.01& 23 57 40.1&    10&    4& 0.000206& 1.13& 1.08& 0.28& 0.12&\nodata \\ 
J034630.1+240507&  3 46 30.16& 24 05 07.1&    22&    5& 0.000417& 0.84& 0.95&-0.18& 0.17&SRS~64413 \\ 
J034630.3+240853&  3 46 30.35& 24 08 53.7&    77&   11& 0.001536& 1.11& 3.27& 0.17& 0.09&\nodata \\ 
J034630.5+235546&  3 46 30.59& 23 55 46.9&    26&    7& 0.000548& 0.82& 1.41& 0.06& 0.10&\nodata \\ 
J034630.8+241124&  3 46 30.85& 24 11 24.1&    17&    5& 0.000340& 0.63& 1.28& 0.06& 0.11&\nodata \\ 
J034631.1+240701&  3 46 31.10& 24 07 02.0&   432&   23& 0.008242& 1.70& 2.02&-0.21& 0.04&HII~1061 \\ 
J034632.0+235858&  3 46 32.04& 23 58 58.1&   861&   34& 0.017366& 3.29& 5.16&-0.17& 0.03&HII~1069 \\ 
J034632.0+240746&  3 46 32.01& 24 07 46.5&     9&    4& 0.000176& 1.02& 1.04& 0.05& 0.21&\nodata \\ 
J034632.4+240547&  3 46 32.42& 24 05 47.7&   112&   12& 0.002108& 1.06& 0.64& 0.17& 0.09&\nodata \\ 
J034634.0+240146&  3 46 34.01& 24 01 46.0&    20&    5& 0.000387& 0.89& 0.66& 0.48& 0.19&\nodata \\ 
J034634.8+240728&  3 46 34.88& 24 07 28.8&    13&    4& 0.000253& 0.65& 0.77& 0.12& 0.20&\nodata \\ 
J034635.4+240134&  3 46 35.43& 24 01 35.0&   632&   29& 0.012193&13.65&19.69& 0.17& 0.03&HHJ~140 \\ 
J034635.7+240754&  3 46 35.73& 24 07 54.5&    42&    8& 0.000798& 0.69& 0.42& 0.21& 0.13&\nodata \\ 
J034635.8+235800&  3 46 35.83& 23 58 00.3&   571&   26& 0.012113& 3.24& 5.68&-0.25& 0.04&HII~1094 \\ 
J034638.7+235804&  3 46 38.78& 23 58 04.2&   119&   12& 0.002394& 0.81& 1.33& 0.12& 0.08&0346388+235805 \\ 
J034639.2+240610&  3 46 39.27& 24 06 10.9&  1093&   34& 0.020173& 1.55& 3.81&-0.49& 0.03&HII~1122 \\ 
J034639.3+240146&  3 46 39.31& 24 01 46.4&   714&   29& 0.013615& 1.32& 0.97&-0.22& 0.04&HII~1124 \\ 
J034640.6+240152&  3 46 40.62& 24 01 52.0&    14&    4& 0.000272& 0.65& 0.49& 0.22& 0.24&SRS~64425 \\ 
J034640.8+240746&  3 46 40.89& 24 07 46.1&    25&    6& 0.000469& 0.95& 1.47& 0.33& 0.16&\nodata \\ 
J034641.3+240426&  3 46 41.33& 24 04 26.4&    67&    9& 0.002001& 1.02& 2.55& 0.28& 0.12&\nodata \\ 
J034641.8+240351&  3 46 41.88& 24 03 51.8&    20&    5& 0.000372& 0.79& 0.82& 0.13& 0.21&\nodata \\ 
J034643.5+235941&  3 46 43.52& 23 59 41.6&   175&   14& 0.003561& 1.33& 1.12&-0.23& 0.07&SRS~62618 \\ 
J034643.6+240120&  3 46 43.65& 24 01 20.5&    18&    5& 0.000343& 0.77& 0.90& 0.28& 0.20&U1125\_01262845 \\ 
J034644.0+235911&  3 46 44.05& 23 59 11.5&    14&    5& 0.000288& 0.66& 1.43& 0.13& 0.18&\nodata \\ 
J034644.1+240218&  3 46 44.19& 24 02 18.6&    20&    5& 0.000375& 0.45& 0.65& 0.16& 0.20&\nodata \\ 
J034644.5+240122&  3 46 44.50& 24 01 22.2&     9&    4& 0.000171& 1.06& 0.82& 0.18& 0.24&\nodata \\ 
J034646.0+235750&  3 46 46.02& 23 57 50.3&    16&    5& 0.000335& 0.77& 0.47& 0.05& 0.15&\nodata \\ 
J034646.6+240757&  3 46 46.65& 24 07 57.6&    50&    8& 0.000955& 1.07& 1.49& 0.13& 0.12&KECK \\ 
J034647.2+240128&  3 46 47.27& 24 01 28.7&   115&   12& 0.002181& 1.07& 2.00& 0.20& 0.09&KECK \\ 
J034647.9+240601&  3 46 48.00& 24 06 01.9&     9&    4& 0.000168& 0.88& 0.90&-0.08& 0.29&\nodata \\ 
J034648.1+240334&  3 46 48.11& 24 03 34.1&    13&    4& 0.000305& 0.73& 0.85& 0.21& 0.27&\nodata \\ 
J034648.5+240039&  3 46 48.59& 24 00 39.5&    27&    6& 0.000519& 0.94& 1.14& 0.34& 0.17&\nodata \\ 
J034649.1+240833&  3 46 49.11& 24 08 33.0&    61&    9& 0.001175& 0.92& 0.60& 0.06& 0.11&KECK \\ 
J034651.3+240615&  3 46 51.38& 24 06 15.8&    43&    7& 0.000803& 1.14& 1.03&-0.25& 0.14&\nodata \\ 
J034651.4+235953&  3 46 51.49& 23 59 54.0&     9&    4& 0.000175& 0.66& 0.95& 0.21& 0.19&\nodata \\ 
J034652.0+240005&  3 46 52.00& 24 00 05.2&    29&    6& 0.000636& 0.96& 1.56& 0.12& 0.14&KECK \\ 
J034652.1+240850&  3 46 52.14& 24 08 50.7&     9&    4& 0.000177& 0.63& 0.52& 0.35& 0.22&\nodata \\ 
J034652.8+240338&  3 46 52.88& 24 03 38.9&    16&    5& 0.000293& 0.57& 0.45&-0.21& 0.20&\nodata \\ 
J034652.9+240037&  3 46 52.99& 24 00 37.6&    17&    5& 0.000571& 0.59& 1.63& 0.21& 0.19&\nodata \\ 
J034653.0+235656&  3 46 53.03& 23 56 56.4&    73&   10& 0.001480& 0.63& 0.65& 0.13& 0.08&KECK \\ 
J034653.6+240325&  3 46 53.63& 24 03 25.2&     7&    3& 0.000129& 0.68& 0.96&-0.15& 0.28&\nodata \\ 
J034653.9+240757&  3 46 53.91& 24 07 57.0&    76&   10& 0.001541& 2.42& 2.93& 0.06& 0.11&MHO~8 \\ 
J034654.4+240642&  3 46 54.40& 24 06 42.6&    11&    4& 0.000213& 1.14& 0.78&-0.16& 0.23&U1125\_01264261 \\ 
J034655.0+240033&  3 46 55.02& 24 00 34.0&    11&    4& 0.000210& 0.83& 0.74& 0.11& 0.19&\nodata \\ 
J034655.3+240024&  3 46 55.39& 24 00 24.5&    11&    4& 0.000210& 0.73& 0.68& 0.11& 0.17&\nodata \\ 
J034655.4+241116&  3 46 55.45& 24 11 16.0&    32&    8& 0.000646& 1.43& 1.90&-0.24& 0.12&MHO~9 \\ 
J034655.6+235622&  3 46 55.70& 23 56 22.8&    20&    6& 0.001552& 0.31& ...&-0.23& 0.16&HHJ~257 \\ 
J034655.7+240151&  3 46 55.76& 24 01 51.5&    28&    6& 0.000525& 0.59& 1.20& 0.24& 0.16&\nodata \\ 
J034657.1+240337&  3 46 57.10& 24 03 37.5&    11&    4& 0.000204& 1.02& 0.82& 0.21& 0.27&\nodata \\ 
J034657.3+240612&  3 46 57.35& 24 06 12.9&     7&    3& 0.000132& 0.52& 0.42& 0.31& 0.28&\nodata \\ 
J034658.1+240140&  3 46 58.19& 24 01 41.0&    60&    9& 0.001134& 1.04& 0.84&-0.37& 0.11&0346582+240141 \\ 
J034659.2+240142&  3 46 59.25& 24 01 42.5&   163&   14& 0.003087& 1.91& 1.63&-0.10& 0.07&HHJ~299 \\ 
J034659.9+240132&  3 46 59.92& 24 01 32.1&    11&    4& 0.000209& 1.02& 0.61&-0.03& 0.18&\nodata \\ 
J034700.7+241322&  3 47 00.76& 24 13 22.3&    46&    9& 0.000964& 1.45& 1.29& 0.08& 0.08&\nodata \\ 
J034701.2+240206&  3 47 01.30& 24 02 06.0&    41&    7& 0.000808& 1.42& 1.59& 0.17& 0.13&\nodata \\ 
J034701.5+241047&  3 47 01.58& 24 10 47.5&    46&    8& 0.000932& 0.90& 0.66& 0.25& 0.10&\nodata \\ 
J034703.5+240934&  3 47 03.54& 24 09 34.7&   433&   24& 0.008656& 1.84& 1.65&-0.28& 0.04&HII~1280 \\ 
J034704.0+235942&  3 47 04.06& 23 59 42.5&    13&    5& 0.000269& 0.53& 0.37&-0.20& 0.16&HII~1284 \\ 
J034704.7+235748&  3 47 04.80& 23 57 48.3&    20&    6& 0.000401& 0.78& 1.07& 0.19& 0.12&\nodata \\ 
J034705.4+240608&  3 47 05.42& 24 06 08.7&    12&    4& 0.000281& 0.57& 0.89& 0.11& 0.19&\nodata \\ 
J034705.9+235943&  3 47 05.91& 23 59 43.5&    47&    8& 0.000924& 0.78& 1.60& 0.21& 0.11&\nodata \\ 
J034707.5+241013&  3 47 07.54& 24 10 13.7&    54&    9& 0.001101& 1.15& 0.93& 0.33& 0.06&KECK \\ 
J034709.1+240307&  3 47 09.12& 24 03 07.6&   376&   21& 0.007586& 3.50& 4.05&-0.19& 0.05&SRS~60765 \\ 
J034709.6+240146&  3 47 09.69& 24 01 46.6&    25&    6& 0.000502& 1.43& 0.97& 0.00& 0.14&0347099+240149 \\ 
J034710.1+240623&  3 47 10.17& 24 06 23.5&    16&    5& 0.000398& 0.53& 0.41& 0.00& 0.17&\nodata \\ 
J034711.2+240647&  3 47 11.23& 24 06 47.7&    14&    5& 0.000278& 0.83& 1.19& 0.26& 0.14&\nodata \\ 
J034711.7+240652&  3 47 11.76& 24 06 52.9&    18&    5& 0.000359& 0.69& 0.74& 0.23& 0.15&\nodata \\ 
J034711.8+241353&  3 47 11.89& 24 13 53.6&    41&   11& 0.000885& 0.93& 0.63&-0.11& 0.09&HHJ~92 \\ 
J034712.5+240112&  3 47 12.53& 24 01 12.3&    11&    4& 0.000223& 0.93& 0.87& 0.09& 0.17&\nodata \\ 
J034712.7+240812&  3 47 12.74& 24 08 12.2&    19&    6& 0.000410& 1.48& 1.20&-0.02& 0.11&\nodata \\ 
J034713.1+240045&  3 47 13.17& 24 00 45.2&    13&    5& 0.000265& 0.74& 1.12&-0.24& 0.11&0347131+240045 \\ 
J034716.5+240741&  3 47 16.53& 24 07 41.9&   277&   19& 0.005657& 1.27& 1.63&-0.40& 0.05&HII~1338 \\ 
J034716.9+241233&  3 47 16.90& 24 12 33.5&    68&   11& 0.003017& 0.72&  ...& 0.12& 0.10&\nodata \\ 
J034718.0+240211&  3 47 18.10& 24 02 11.0&  1158&   37& 0.023297& 2.75& 4.70&-0.15& 0.03&HII~1355 \\ 
J034722.6+240010&  3 47 22.65& 24 00 10.2&    28&    7& 0.000579& 1.02& 0.75& 0.10& 0.08&\nodata \\ 
J034725.3+240255&  3 47 25.31& 24 02 55.8&   118&   14& 0.002442& 1.16& 0.73&-0.21& 0.06&HHJ~427 \\ 
\enddata 
\tablenotetext{a}{Optical/IR name associated with each X-ray source.  A blank 
        entry denotes an AGN candidate.} 
\end{deluxetable} 

\clearpage 
\begin{deluxetable}{lr} 
\tablewidth{0pt} 
\tabletypesize{\scriptsize} 
\tablecaption{Position uncertainties.} 
\tablehead{ 
\colhead{Catalog}   & \colhead{uncertainty} \\ 
\colhead{}          & \colhead{($\arcsec$)} } 
\startdata 
\citet{Her47}\tablenotemark{a} & 0.5\\ 
\citet{Har82} &30.0\\ 
\citet{Ham93} &10.0\\ 
\citet{Sta94} &30.0\\ 
\citet{Mic96} &30.0\\ 
\citet{Sta98} & 1.5\\ 
\citet{Mic99} &15.0\\ 
\citet{Pin00} & 1.5\\ 
\citet{Mar00} & 1.0\\ 
\citet{Kri01} & 1.0\\ 
2MASS 2001    & 1.0\\ 
USNO-A 2001    & 0.5\\ 
\enddata 
\tablenotetext{a}{Accurate positions for this catalog were derived by 
        \citet{Eic70}.} 
\end{deluxetable} 

\clearpage 
\begin{deluxetable}{llllllllll} 
\rotate 
\tablewidth{0pt} 
\tablecaption{Names for the previously cataloged {\it Chandra} detections.} 
\tabletypesize{\scriptsize} 
\tablehead{ 
\colhead{CXOP}                    & \colhead{HII/}                     & 
\colhead{SCG94/}                  & \colhead{MHO}                      & 
\colhead{BPL}                     & \colhead{KECK}                     & 
\colhead{2MASS}                   & \colhead{USNO-A}                   &  
\colhead{Other}                   & \colhead{$\Delta$\tablenotemark{a}}\\ 
\colhead{}                        & \colhead{HHJ}                      & 
\colhead{MSK}                     & \colhead{}                         & 
\colhead{}                        & \colhead{}                         & 
\colhead{}                        & \colhead{}                         & 
\colhead{}                        & \colhead{($\arcsec$)}              } 
\startdata 
J034604.5+240957 & \nodata&     \nodata & \nodata& BPL~88 & \nodata&       \nodata&         \nodata&   \nodata&   \nodata\\
J034612.5+240347 & \nodata&    120 / 84 & \nodata& \nodata& \nodata&       \nodata&         \nodata&   \nodata&   \nodata\\
J034612.7+240314 & HII~930&    120 / 84 & \nodata& \nodata& \nodata&        \nodata& U1125\_01258659& SRS~66387&      3.57\\
J034615.8+241122 & HII~956&    122 / 86 & \nodata& \nodata& \nodata&        \nodata& U1125\_01259068&  HD~23479&      2.38\\
J034619.4+235628 & \nodata& \nodata/ 88 & \nodata& \nodata& \nodata&       \nodata&         \nodata&   \nodata&   \nodata\\
J034619.5+235653 & HII~980&    123 / 88 & \nodata& \nodata& \nodata& 0346195+235653& U1125\_01259575& Merope/HD~23480/23~Tau& 0.93\\
J034620.1+240028 & HII~989&     \nodata & \nodata& \nodata& \nodata&       \nodata&         \nodata&   \nodata&   \nodata\\
J034620.8+235858 & \nodata& 125 /\nodata& \nodata& \nodata& \nodata&       \nodata&         \nodata&   \nodata&   \nodata\\
J034623.3+240151 & HHJ~195& 128 /\nodata& \nodata& \nodata& \nodata& 0346234+240151&         \nodata&   \nodata&      1.17\\
J034623.6+235538 & \nodata& 129 /\nodata& \nodata& \nodata& \nodata& 0346237+235540&         \nodata&   \nodata&      2.51\\
J034625.2+240936 & HHJ~429&    130 / 90 & \nodata& \nodata& \nodata& 0346253+240936& U1125\_01260355& HCG~224/SRS~64388& 1.05\\
J034627.9+235532 & \nodata&     \nodata & \nodata& \nodata& \nodata& 0346279+235525&         \nodata&   \nodata&      6.44\\
J034628.8+235820 & \nodata&     \nodata & \nodata& \nodata& \nodata& 0346290+235822&         \nodata&   \nodata&      2.42\\
J034629.1+240839 & \nodata&     \nodata & \nodata& \nodata& \nodata& 0346290+240839&         \nodata&   \nodata&      2.49\\
J034630.1+240507 & \nodata& 135 /\nodata& \nodata& \nodata& \nodata& 0346302+240507&         \nodata& SRS~64413&      1.07\\
J034631.1+240701 &HII~1061&    136 / 94 & \nodata& \nodata& \nodata& 0346311+240702& U1125\_01261151&HCG~224/SRS~64400& 1.07\\
J034632.0+235858 &HII~1069& 137 /\nodata& \nodata& \nodata& \nodata& 0346321+235858&         \nodata&   HCG~225&      1.21\\
J034632.4+240547 & \nodata& 138 /\nodata& \nodata& \nodata& \nodata&       \nodata&         \nodata&   \nodata&   \nodata\\
J034635.4+240134 & HHJ~140&     \nodata & \nodata& \nodata& \nodata& 0346355+240135&         \nodata&   \nodata&      1.32\\
J034635.8+235800 &HII~1094&    141 / 97 & \nodata& \nodata& \nodata& 0346358+235801&         \nodata&HCG~227/SRS~64468& 1.38\\
J034638.7+235804 & \nodata&      \nodata& \nodata& \nodata& \nodata& 0346388+235805& U1125\_01262204&   \nodata&      1.04\\
J034639.2+240610 &HII~1122&   144 / 103 & \nodata& \nodata& \nodata& 0346393+240611& U1125\_01262280&HD~23511/SRS~64407& 1.00\\
J034639.3+240146 &HII~1124&   146 / 104 & \nodata& \nodata& \nodata& 0346393+240146& U1125\_01262279&   \nodata&      1.02\\
J034640.6+240152 & \nodata&   146 / 104 & \nodata& \nodata& \nodata&       \nodata&         \nodata& SRS~64425&   \nodata\\
J034641.3+240426 & \nodata& 148 /\nodata& \nodata& \nodata& \nodata&       \nodata&         \nodata&   \nodata&   \nodata\\
J034643.5+235941 & \nodata&   149 / 107 & \nodata& \nodata& \nodata& 0346435+235942& U1125\_01262827& SRS~62618&      1.29\\
J034643.6+240120 & \nodata&      \nodata& \nodata& \nodata& \nodata&        \nodata& U1125\_01262845&   \nodata&      0.82\\
J034644.0+235911 & \nodata& \nodata/ 107& \nodata& \nodata& \nodata&       \nodata&         \nodata&   \nodata&   \nodata\\
J034646.6+240757 & \nodata&      \nodata& \nodata& \nodata&     Yes&       \nodata&         \nodata&   \nodata&      1.38\\
J034647.2+240128 & \nodata& 151 /\nodata& \nodata& \nodata&     Yes&       \nodata&         \nodata&   \nodata&      1.67\\
J034647.9+240601 & \nodata& 150 /\nodata& \nodata& \nodata& \nodata&       \nodata&         \nodata&   \nodata&   \nodata\\
J034649.1+240833 & \nodata&      \nodata& \nodata& \nodata&     Yes&       \nodata&         \nodata&   \nodata&      1.47\\
J034651.3+240615 & \nodata&      \nodata& \nodata& \nodata&     Yes&       \nodata&         \nodata&   \nodata&      1.54\\
J034652.0+240005 & \nodata&      \nodata& \nodata& \nodata&     Yes&       \nodata&         \nodata&   \nodata&      1.55\\
J034653.0+235656 & \nodata& 155 /\nodata& \nodata& \nodata&     Yes&       \nodata&         \nodata&   \nodata&      2.43\\
J034653.9+240757 & \nodata&      \nodata&  MHO~8 & \nodata&     Yes& 0346539+240757&         \nodata&   \nodata&      0.98\\
J034654.4+240642 & \nodata&      \nodata& \nodata& \nodata& \nodata&        \nodata& U1125\_01264261&   \nodata&      2.15\\
J034655.4+241116 & \nodata&      \nodata&  MHO~9 & BPL~116& \nodata& 0346553+241116&         \nodata&   \nodata&      1.93\\
J034655.6+235622 & HHJ~257& 155 /\nodata& \nodata& \nodata& \nodata&       \nodata&         \nodata&   \nodata&   \nodata\\
J034658.1+240140 & \nodata& 157 /\nodata& \nodata& \nodata& \nodata& 0346582+240141&         \nodata&   \nodata&      1.01\\
J034659.2+240142 & HHJ~299& 157 /\nodata& \nodata& \nodata& \nodata& 0346593+240142& U1125\_01264960&   \nodata&      0.92\\
J034659.9+240132 & \nodata& 157 /\nodata& \nodata& \nodata& \nodata&       \nodata&         \nodata&   \nodata&   \nodata\\
J034703.5+240934 &HII~1280&   160 / 117 & \nodata& \nodata&     Yes& 0347035+240935& U1125\_01265582&HCG~249/SRS~62537& 1.55\\
J034704.0+235942 &HII~1284& \nodata/ 118& \nodata& \nodata& \nodata& 0347042+235942& U1125\_01265693&HD~23585/SRS~62623& 2.08\\
J034705.9+235943 & \nodata& \nodata/ 118& \nodata& \nodata& \nodata&       \nodata&         \nodata&   \nodata&   \nodata\\
J034707.5+241013 & \nodata& 163 /\nodata& \nodata& \nodata&     Yes&       \nodata&         \nodata&   \nodata&      4.75\\
J034709.1+240307 & \nodata& 166 /\nodata& \nodata& \nodata& \nodata& 0347091+240307&         \nodata& SRS~60765&      0.65\\
J034709.6+240146 & \nodata&      \nodata& \nodata& \nodata& \nodata& 0347099+240149&         \nodata&   \nodata&      4.29\\
J034711.8+241353 & HHJ~92 &      \nodata& HHJ~92 & BPL~131& \nodata& 0347118+241354&         \nodata&   \nodata&      0.65\\
J034713.1+240045 & \nodata&      \nodata& \nodata& \nodata& \nodata& 0347131+240045& U1125\_01266937&   \nodata&      0.24\\
J034716.5+240741 &HII~1338&   170 / 128 & \nodata& \nodata& \nodata& 0347165+240742&         \nodata&  HD~23608&      0.39\\
J034716.9+241233 & \nodata& 171 /\nodata& \nodata& \nodata& \nodata&       \nodata&         \nodata&   \nodata&   \nodata\\
J034718.0+240211 &HII~1355&   172 / 130 & \nodata& \nodata& \nodata& 0347181+240211&         \nodata&HCG~262/SRS~60774& 0.58\\
J034725.3+240255 & HHJ~427& 179 /\nodata& \nodata& \nodata& \nodata& 0347253+240257&         \nodata&   \nodata&      1.25\\
\enddata 
\tablecomments{HII: \citet{Her47} catalog.  
        HHJ: \citet{Ham93} catalog. 
        SCG94: {\it ROSAT} catalog in \citet{Sta94}.  
        MSK: {\it ROSAT} catalog in \citet{Mic96}. 
        MHO: Mount Hopkins Observatory sources in \citet{Sta98}. 
        BPL: Burrell Pleiades candidates as designated 
        by \citet{Pin00}. 
        KECK: Sources detected by Eduardo Mart\'{\i}n with the 
        Keck II Telescope (see Paper I). 
        2MASS: Second Incremental Release of the Point Source 
        Catalog. 
        USNO-A: Catalog A from the US Naval Observatories. 
        Other: Additional references given by the SIMBAD 
        database.} 
\tablenotetext{a}{Offset from optical or IR counterparts from Keck 
        astrometry \citep{Kri01}, 2MASS (2MASS 2001), or USNO 
        (USNO-A 2001), in order of preference.} 
\end{deluxetable}  

\clearpage 
\begin{deluxetable}{llrrrrrrrrcl} 
\tablewidth{0pt} 
\tabletypesize{\scriptsize} 
\tablecaption{IR and optical data for {\it Chandra} sources.} 
\tablehead{               
\colhead{CXOP}                & \colhead{Name}          & 
\colhead{$\log{L_{\rm X}}$}   & \colhead{B}             & 
\colhead{V}                   & \colhead{R}             & 
\colhead{I}                   & \colhead{J}             & 
\colhead{H}                   & \colhead{K}             & 
\colhead{MK}                  & \colhead{Status\tablenotemark{a}}\\ 
\colhead{}                    & \colhead{}              & 
\colhead{(erg~s$^{-1}$)}      & \colhead{}              & 
\colhead{}                    & \colhead{}              & 
\colhead{}                    & \colhead{}              & 
\colhead{}                    & \colhead{}              & 
\colhead{}                    & \colhead{}              } 
\startdata 
J034604.5+240957& BPL~88         & \nodata& \nodata& \nodata& \nodata&  15.74 & \nodata& \nodata& \nodata& \nodata& NM\\
J034612.7+240314& HII~930        &  29.08 &  15.30 &  14.08 &  14.20 &  12.60 & \nodata& \nodata& \nodata&  K~V   &  PM\\
J034615.8+241122& HII~956        &  29.31 &   8.12 &   7.84 &   8.30 & \nodata& \nodata& \nodata& \nodata&  A7~V + F6 &  PM\\
J034619.5+235653& HII~980        &  29.6  &   3.96 &   4.06 &   4.20 & \nodata& \nodata& \nodata& \nodata&  B6~IV + G&  PM\\
J034623.3+240151& HHJ~195        &  28.13 & \nodata&  18.5  &  17.50 &  15.10 & 13.71 &  13.07 &  12.77 & \nodata&  PC\\
J034623.6+235538& 0346237+235540 & \nodata& \nodata& \nodata& \nodata& \nodata&  16.98 &  16.10 &  15.22 & \nodata& AGN\\
J034625.2+240936& HHJ~429        &  29.19 &  18.00 &  16.00 &  14.70 &  13.20 & 12.01 &  11.36 &  11.11 &  M     &  PM\\
J034627.9+235532& 0346279+235525 & \nodata& \nodata& \nodata& \nodata& \nodata&  14.79 &  14.30 &  14.03 & \nodata& AGN\\
J034628.8+235820& 0346290+235822 & \nodata& \nodata& \nodata& \nodata& \nodata&  13.06 &  12.65 &  12.48 & \nodata& NM\\
J034629.1+240839& 0346290+240839 & \nodata& \nodata& \nodata& \nodata& \nodata&  16.00 &  15.13 &  14.85 & \nodata& AGN\\
J034630.1+240507& SRS~64413      & \nodata&  17.06 &  16.07 & \nodata& \nodata& 12.87 &  12.28 &  12.06 & \nodata& NM\\
J034631.1+240701& HII~1061       &  29.06 &  14.64 &  13.49 &  13.30 &  12.26 & 11.27 &  10.60 &  10.44 &  K5~V  &  PM\\
J034632.0+235858& HII~1069       & \nodata&  14.30 &  14.55 & \nodata& \nodata& 10.55 &   9.88 &   9.69 &  K3~V  & NM\\
J034635.4+240134& HHJ~140        &  29.48 & \nodata&  19.00 &  17.80 &  15.40 & 13.68 &  13.11 &  12.83 &  M~V   &  PC\\
J034635.8+235800& HII~1094       &  29.19 &  15.26 &  13.90 & \nodata&  12.34 & 11.50 &  10.84 &  10.66 &  K~V   &  PM\\
J034638.7+235804& 0346388+235805 & \nodata&  17.30 & \nodata&  15.80 & \nodata& 14.25 &  13.69 &  13.50 & \nodata& NM\\
J034639.2+240610& HII~1122       &  29.06 &   9.59 &   9.17 &   9.70 & \nodata&   8.43 &   8.22 &   8.18 &  F4~V + K &  PM\\
J034639.3+240146& HII~1124       &  29.27 &  13.14 &  12.20 &  12.50 &  11.39 & 10.45 &   9.97 &   9.85 &  K3~V  &  PM\\
J034640.6+240152& SRS~64425      &  27.85 &  13.21 &  12.27 & \nodata& \nodata& \nodata& \nodata& \nodata&  K2~V  &  PM\\
J034643.5+235941& SRS~62618      &  28.67 &  17.63 &  16.12 &  15.10 & \nodata& 12.45 &  11.85 &  11.54 & \nodata&  PC\\
J034643.6+240120& U1125\_01262845& \nodata&  19.70 & \nodata&  17.30 & \nodata& \nodata& \nodata& \nodata& \nodata& NM\\
J034646.6+240757& KECK           & \nodata& \nodata& \nodata&  22.39 &  22.19 & \nodata& \nodata& \nodata& \nodata& AGN\\
J034647.2+240128& KECK           & \nodata& \nodata& \nodata&  21.64 &  21.21 & \nodata& \nodata& \nodata& \nodata& AGN\\
J034649.1+240833& KECK           & \nodata& \nodata& \nodata&  22.07 &  21.72 & \nodata& \nodata& \nodata& \nodata& AGN\\
J034652.0+240005& KECK           & \nodata& \nodata& \nodata&  20.66 &  20.67 & \nodata& \nodata& \nodata& \nodata& AGN\\
J034653.0+235656& KECK           & \nodata& \nodata& \nodata&  20.29 &  19.84 & \nodata& \nodata& \nodata& \nodata& AGN\\
J034653.9+240757& MHO~8          &  28.52 & \nodata&  18.92 & \nodata&  15.76 & 14.10 &  13.48 &  13.20 &  M~V   &  PM\\
J034654.4+240642& U1125\_01264261& \nodata&  17.40 & \nodata&  14.30 & \nodata& \nodata& \nodata& \nodata& \nodata& NM\\
J034655.4+241116& MHO~9          &  27.92 & \nodata&  19.02 & \nodata&  15.55 & 13.78 &  13.22 &  12.89 &  M~V   &  PM\\
J034655.6+235622& HHJ~257        &  28.32 & \nodata&  18.40 &  17.00 &  14.70 & \nodata& \nodata& \nodata& \nodata&  PC\\
J034658.1+240140& 0346582+240141 & \nodata& \nodata& \nodata& \nodata& \nodata&  13.95 &  13.27 &  12.95 &  M~V   & NM\\
J034659.2+240142& HHJ~299        &  28.72 &  19.90 &  17.60 &  16.60 &  14.40 & 13.00 &  12.38 &  12.12 &  M~V   &  PM\\
J034703.5+240934& HII~1280       &  29.01 &  15.85 &  14.45 &  13.80 &  12.84 & 11.63 &  10.96 &  10.76 &  K7.5  &  PM\\
J034704.0+235942& HII~1284       &  27.58 &   8.51 &   8.25 &   8.70 & \nodata&   7.76 &   7.68 &   7.63 &  A9~V + K &  PM\\
J034707.5+241013& KECK           & \nodata& \nodata& \nodata&  23.80 &  23.49 & \nodata& \nodata& \nodata& \nodata& AGN\\
J034709.1+240307& SRS~60765      &  29.03 &  18.33 &  16.52 & \nodata& \nodata& 12.59 &  11.94 &  11.73 & \nodata&  PC\\
J034709.6+240146& 0347099+240149 & \nodata& \nodata& \nodata& \nodata& \nodata&  16.39 &  15.63 &  15.16 & \nodata& AGN\\
J034711.8+241353& HHJ~92         &  28.17 & \nodata&  19.52 &  18.60 &  15.87 & 14.03 &  13.51 &  13.20 & \nodata&  PM\\
J034713.1+240045& 0347131+240045 & \nodata&  18.60 & \nodata&  19.10 & \nodata& 14.88 &  14.18 &  13.91 & \nodata& NM\\
J034716.5+240741& HII~1338       &  28.67 &   8.99 &   8.57 & \nodata& \nodata&   7.74 &   7.58 &   7.50 &  F3~V + F6  &  PM\\
J034718.0+240211& HII~1355       &  29.56 &  15.26 &  13.90 & \nodata&  12.25 & 11.05 &  10.36 &  10.19 &  K6n   &  PM\\
J034725.3+240255& HHJ~427        &  28.53 & \nodata&  17.60 &  15.20 &  14.60 & 12.65 &  12.03 &  11.79 &  M~V   &  PM\\
\enddata 
\tablenotetext{a}{Pleiades members (PM) and Pleiades candidates (PC) are 
        determined using the proper motion probability and photometric 
        proterties as listed by \citet{Sta98}, \citet{Bel98}, 
        \citet{Mic96}, \citet{Sch95}, and \citet{Ham93} ($\S3.1.1$).  
        Non-members (NM) are sources where $\log{(f_{x}/f_{v})}\leq-1.0$ 
        that do not show any of the characteristics of a Pleiades star. 
        AGN candidates (AGN) have $-1.0\leq\log{(f_{x}/f_{v})}\leq1.2$.} 
\end{deluxetable} 

\clearpage 
\begin{deluxetable}{llrrrrrrrc} 
\tablewidth{0pt} 
\tabletypesize{\scriptsize} 
\tablecaption{Pleiades members not detected by {\it Chandra}.} 
\tablehead{ 
\colhead{Name}                 &\colhead{$\log{L_{\rm X}}$}   & 
\colhead{B}                    & \colhead{V}                  & 
\colhead{R}                    & \colhead{I}                  & 
\colhead{J}                    & \colhead{H}                  & 
\colhead{K}                    & \colhead{MK}                 \\ 
\colhead{}                     &\colhead{(erg~s$^{-1}$)}      & 
\colhead{}                     & \colhead{}                   & 
\colhead{}                     & \colhead{}                   & 
\colhead{}                     & \colhead{}                   & 
\colhead{}                     & \colhead{}                   } 
\startdata 
MHO~10   & $<27.84$ & ...&20.18&  ...&16.87&15.00&14.41&14.06&  ...   \\ 
MHO~11   & $<28.17$ & ...&21.44&  ...&16.98&14.84&14.20&13.99&  ...   \\ 
HCG~254  & $<28.17$ & ...&19.34&18.60&16.00&14.26&13.72&13.45&  K0~V  \\ 
 HII~1362& $<27.87$ &8.50& 8.25&  ...&  ...& 7.73& 7.68& 7.63&  A7    \\ 
 HII~1375& $<27.98$ &6.30& 6.29&  ...&  ...& 6.21& 6.28& 6.27&  A0~V + A~SB \\ 
\enddata 
\end{deluxetable} 

\clearpage 
\begin{deluxetable}{lcllccllllll} 
\rotate 
\tablewidth{0pt} 
\tabletypesize{\scriptsize} 
\tablecaption{Spectral information for six X-ray bright Pleiades members.} 
\tablehead{ 
\colhead{Name}                  & \colhead{MK}                   & 
\colhead{kT$_1$}                & \colhead{EM$_1$}               & 
\colhead{kT$_2$}                & \colhead{EM$_2$}               & 
\colhead{${\rm Ne}$}            & \colhead{${\rm Mg}$}           & 
\colhead{${\rm Si}$}            & \colhead{${\rm Fe}$}           & 
\colhead{$\chi^2$}              & \colhead{Date}                 \\ 
\colhead{}                      & \colhead{}                     & 
\colhead{(keV)}                 & \colhead{($10^{52}$cm$^{-3}$)} & 
\colhead{(keV)}                 & \colhead{($10^{52}$cm$^{-3}$)} & 
\colhead{}                      & \colhead{}                     & 
\colhead{}                      & \colhead{}                     & 
\colhead{}                      & \colhead{}                     } 
\startdata 
HII~1094 & K~V   & $0.48\pm0.04$ &$0.7\pm0.2$&\nodata& \nodata& $1.00$        & 1.00          & $1.00$        & $0.25\pm0.08$ &0.90 &1999 Sep 18\\
HII~1094 & K~V   & $0.62\pm0.06$ &$0.7\pm0.1$&   3.5 &     0.4& $2.50\pm0.45$ & 1.00          & $2.00\pm0.50$ & $0.52\pm0.14$ &0.75 &2000 Mar 20\\
HII~1124 & K3~V  & $0.34\pm0.02$ &$1.1\pm0.2$&   3.5 &     0.4& $1.00$        & 1.00          & $1.00$        & $0.23\pm0.07$ &0.84 &1999 Sep 18\\
HII~1124 & K3~V  & $0.66\pm0.05$ &$1.3\pm0.3$&\nodata& \nodata& $1.00$        & 1.00          & $1.00$        & $0.19\pm0.06$ &0.77 &2000 Mar 20\\
HII~1355 & K6    & $0.33\pm0.02$ &$0.9\pm0.1$&   3.5 &     1.3& $2.59\pm0.32$ & 1.00          & $3.00\pm1.00$ & $0.55\pm0.15$ &0.83 &1999 Sep 18\\
HII~1355 & K6    & $0.40\pm0.05$ &$1.0\pm0.6$&   3.5 &     0.8& $1.83\pm0.61$ & 1.00          & $1.00$        & $0.16\pm0.12$ &0.84 &2000 Mar 20\\
HII~1122 & F4~V + K  & $0.45\pm0.04$ &$0.6\pm0.1$&\nodata& \nodata& $1.00$       & $2.79\pm0.74$ & 1.00          & $1.00$        &1.09 &2000 Mar 20\\
HII~956  & A7~V + F6 & $0.57\pm0.01$ &$1.4\pm0.1$&\nodata& \nodata& $1.50\pm0.30$ & $1.50\pm0.30$ & 1.00          & $1.00$        &1.61 &1999 Sep 18\\
HII~980  & B6~IV + G & $0.53\pm0.02$ &$2.4\pm0.2$&\nodata& \nodata& $1.50\pm0.30$ & $1.50\pm0.20$ & $1.50\pm0.20$ & $0.63\pm0.05$ &0.93 &1999 Sep 18\\
HII~980  & B6~IV + G & $0.59\pm0.02$ &$1.6\pm0.6$&\nodata& \nodata& $1.00$       & $1.39\pm0.50$ & $1.69\pm0.74$ & $0.69\pm0.17$ &1.05 &2000 Mar 20\\
\enddata 
\end{deluxetable} 

\clearpage 
\begin{deluxetable}{llcrrrrrrl} 
\tablewidth{0pt} 
\tabletypesize{\scriptsize} 
\tablecaption{X-ray properties of Pleiades members in the {\it Chandra} FOV.} 
\tablehead{ 
\colhead{Name}              & \colhead{CXOP}              & 
\colhead{MK}                & \colhead{V}                 & 
\colhead{B$-$V}             & \colhead{V$-$I}             & 
\colhead{ $^{\log{L_{\rm X}}}_{\rm (ergs~s^{-1})}$ }      & 
\colhead{HR}                & \colhead{KS$_{s}$}          & 
\colhead{Variability}       } 
\startdata 
\cutinhead{Inactive star sample} 
HII~1375   & \nodata          &  A0~V + A~SB &  6.29 &    0.01 &  \nodata &$<$27.98& \nodata&   \nodata & \nodata\\ 
HII~1284   & J034704.0+235942 &  A9~V + K &  8.25 &    0.26 &  \nodata &   27.58&   -0.20&     0.53 & No\\ 
HII~1362   & \nodata          &  A7   &  8.25 &    0.25 &  \nodata &$<$27.87&  \nodata&  \nodata & \nodata\\ 
SRS~64425  & J034640.6+240152 & K2~V  & 12.27 &    0.94 &  \nodata &   27.85&    0.22&     0.65 & No\\ 
HCG~254    & \nodata          &\nodata& 19.34 & \nodata &     3.34 &$<$28.17& \nodata&  \nodata & \nodata\\ 
HHJ~92     & J034711.8+241353 &\nodata& 19.52 & \nodata &     3.65 &   28.17&   -0.11&     0.93 & No\\ 
MHO~10     & \nodata          & M     & 20.18 & \nodata &     3.31 &$<$27.84& \nodata&    \nodata& \nodata\\ 
MHO~11     & \nodata          & M     & 21.44 & \nodata &     4.46 &$<$28.17& \nodata&    \nodata& \nodata\\ 
\cutinhead{Active K star sample} 
HII~1124   & J034639.3+240146 &  K3~V & 12.20 &    0.94 &     0.81 &   29.27&   -0.22&      1.32& Yes\\ 
HII~1061   & J034631.1+240701 & K5~V  & 13.49 &    1.15 &     1.23 &   29.06&   -0.21&      1.70& flaring\\ 
HII~1355   & J034718.0+240211 &  K6   & 13.90 &    1.36 &     1.65 &   29.56&   -0.15&      2.75& flaring\\ 
HII~1094   & J034635.8+235800 &  K~V  & 13.90 &    1.36 &     1.56 &   29.19&   -0.25&      3.24& flaring\\ 
HII~930    & J034612.7+240314 &  K~V  & 14.08 &    1.22 &     1.48 &   29.08&   -0.19&      1.59& flaring\\ 
HII~1280   & J034703.5+240934 &  K7.5 & 14.45 &    1.40 &     1.61 &   29.01&   -0.28&      1.84& flaring\\ 
\cutinhead{Active B6-F4 star sample} 
HII~980    & J034619.5+235653 & B6~IV + G&  4.06 &   -0.10 &  \nodata &   29.60&   -0.34&      1.91 & Yes\\ 
HII~956    & J034615.8+241122 & A7~V + F6  &  7.84 &    0.28 &  \nodata &   29.31&   -0.48&      1.91 & Yes\\ 
HII~1338   & J034716.5+240741 &  F3~V + F6 &  8.57 &    0.42 &  \nodata &   28.67&   -0.40&      1.27 & Yes\\ 
HII~1122   & J034639.2+240610 &  F4~V + K&  9.17 &    0.42 &  \nodata &   29.06&   -0.49&      1.55 & Yes\\ 
\cutinhead{Active M star sample} 
HHJ~429    & J034625.2+240936 & M     & 16.00 &    2.00 &     2.80 &   29.19&   -0.11&       2.53& flaring\\ 
HHJ~299    & J034659.2+240142 &  M~V  & 17.60 &    2.30 &     3.20 &   28.72&   -0.10&       1.91& flaring\\ 
HHJ~427    & J034725.3+240255 &  M~V  & 17.60 & \nodata &     3.00 &   28.53&   -0.21&       1.16& flaring\\ 
MHO~8      & J034653.9+240757 & M~V   & 18.92 & \nodata &     3.16 &   28.52&    0.06&      2.42 & flaring\\ 
MHO~9      & J034655.4+241116 & M~V   & 19.02 & \nodata &     3.47 &   27.92&   -0.24&      1.43 & Yes\\ 
\cutinhead{Possible Members\tablenotemark{a}} 
SRS~62618  & J034643.5+235941 &\nodata& 16.12 &    1.51 &  \nodata &   28.67&   -0.23&       1.33& flaring\\ 
SRS~60765  & J034709.1+240307 &\nodata& 16.52 &    1.81 &  \nodata &   29.03&   -0.19&       3.50& flaring\\ 
HHJ~257    & J034655.6+235622 &\nodata& 18.40 & \nodata &     3.70 &   28.32&   -0.23&      0.31 & No\\ 
HHJ~195    & J034623.3+240151 &\nodata& 18.50 & \nodata &     3.40 &   28.13&   -0.24&      1.52 & Yes\\ 
HHJ~140    & J034635.4+240134 & M~V   & 19.00 & \nodata &     3.60 &   29.48&    0.17&     13.65 & flaring\\ 
\enddata 
\tablenotetext{a}{X-ray luminosities are computed assuming that the stars are 
        located at the distance of the Pleiades (127~pc).} 
\end{deluxetable} 

\end{document}